\documentclass[aps,twocolumn,prb,amsmath,amssymb,floatfix,reprint]{revtex4-2} 

\usepackage[utf8]{inputenc}
\usepackage{graphicx}
\usepackage{bm}
\usepackage{color}
\usepackage[colorlinks=true,citecolor=blue]{hyperref}
\usepackage{graphicx}
\usepackage{braket}
\usepackage{physics} 
\usepackage{comment} 
\makeatletter
\AtBeginDocument{\let\LS@rot\@undefined}
\makeatother

\newcommand{\innerp}[2]{\left\langle #1 \vert #2 \right\rangle}

\begin{document}

\title{Entanglement measures of Majorana bound states}
\author{Vimalesh Kumar Vimal}

\author{Jorge Cayao}
\affiliation{Department of Physics and Astronomy, Uppsala University, Box 516, S-751 20 Uppsala, Sweden}
\date{\today}

\begin{abstract}
Majorana bound states emerge in topological superconductors as zero-energy edge states exhibiting spatial nonlocality. Despite the enormous advances, the detection of Majorana bound states is still challenging, mainly because topologically trivial Andreev bound states produce similar signatures.  In this work, we consider a topological superconductor with Majorana bound states coupled to quantum dots and investigate the dynamics of their quantum correlations with the aim to explore their entanglement properties. In particular, we characterize entanglement by using concurrence and discord, which are also complemented by entanglement dynamics and return probability. We find that Majorana bound states at truly zero energy can transform an initially entangled system into its classical state, while they can create maximally entangled states at a finite energy overlap. Interestingly, we show that the system can generate  a maximally entangled state between MBSs and a quantum dot by simply controlling the Majorana nonlocality.  We demonstrate that these results hold in scenarios when the initial state is either maximally entangled or separable, albeit in the latter, maximally entangled states are achieved in the long time dynamics.  Furthermore, we contrast our findings with those produced by a regular fermion and obtain very distinct entanglement signatures. Our work offers an alternative approach to characterizing Majorana bound states, which can also be useful towards their utilization for quantum information tasks.
\end{abstract}
\maketitle


\section{Introduction}
Majorana bound states (MBSs) have become one of the central topics in condensed matter physics \cite{tanaka2011symmetry,sato2017topological,Aguadoreview17,lutchyn2018majorana,frolov2019quest,zhang2019next,prada2019andreev,flensberg2021engineered,tanaka2024theory} largely due to their promising properties for fault tolerant quantum computation \cite{sarma2015majorana,Lahtinen_2017,beenakker2019search,aguado2020majorana,Marra_2022}.  MBSs were initially predicted to appear in the topological phase of spinless $p$-wave superconductors, which later were shown to be realized by combining spin-orbit coupling, an external magnetic field, and conventional spin-singlet $s$-wave superconductivity; see, e.g., Ref.\,\cite{tanaka2024theory}. MBSs emerge at zero energy with their wavefunctions located at the system edges, thus exhibiting an inherent spatial nonlocality. In this regard, truly zero-energy MBSs are needed for realizing qubits that are robust against local perturbations \cite{sarma2015majorana}.
Moreover, zero-energy signatures can be also produced by trivial Andreev bound states (ABSs) \cite{Bagrets:PRL12,Pikulin2012A,PhysRevB.86.100503,PhysRevB.86.180503,PhysRevB.91.024514,PhysRevB.98.245407,PhysRevLett.123.117001,PhysRevB.100.220502,PhysRevB.102.245431,DasSarma2021Disorder,PhysRevB.104.134507,PhysRevB.107.184509,PhysRevB.107.184519,PhysRevB.108.035401, PhysRevB.110.085414}, which proliferate in real devices and have challenged the detection of MBSs \cite{prada2019andreev}.

A less explored property of MBSs is their spatial nonlocality, which can be revealed by inducing a finite spatial overlap between Majorana wavefunctions \cite{PhysRevB.86.180503,PhysRevB.86.220506,PhysRevB.87.024515,PhysRevB.96.205425}. In this case, MBSs acquire finite energy splitting, which, although not beneficial for realizing topological qubits, is useful for distinguishing the inherent Majorana nature. Following this idea, it has been recently shown that testing the Majorana nonlocality allows to distinguish between MBSs and trivial Andreev bound states \cite{PhysRevB.92.075443,cayao2018finite,PhysRevB.98.085125,PhysRevB.104.L020501,baldo2023zero,PhysRevLett.130.116202,PhysRevB.106.L100502,PhysRevResearch.6.L012062,PhysRevB.105.L161403}. Majorana nonlocality has also been explored by interference effects when coupling  topological superconductors and  quantum dots (QDs) \cite{PhysRevLett.106.090503, PhysRevB.84.201308, PhysRevB.98.085125}, with shot noise signatures signaling the nonlocal nature of MBSs \cite{PhysRevB.105.205430}. This interference effect has also been shown to provide a way to distinguish between MBSs and Andreev bound states by exploiting  the Josephson effect \cite{PhysRevB.106.L100502,PhysRevResearch.6.L012062}, nonlocal conductance \cite{PhysRevB.106.L100502,drechsler2024electroninterferenceprobemajorana}, and also finite frequency noise \cite{PhysRevB.109.195410}; see also Refs.\,\cite{PhysRevB.91.081405, PhysRevB.92.205422, PhysRevLett.124.096801, PhysRevB.107.155416, PhysRevB.108.L121407} for noise signatures of MBSs. Furthermore, the Majorana nonlocality can be detected  by the Majorana fractional entropy \cite{PhysRevB.92.195312,PhysRevLett.123.147702,PhysRevB.103.075440,PhysRevB.104.205406}, which can be of particular relevance for experimental measurements of  entropy in nanosystems \cite{Hartman2018,Kleeorin2019,e23060640,e24030417,PhysRevLett.128.146803,PhysRevLett.129.227702}. The Majorana nonlocaly thus offers a solid way for the unambiguous detection of MBSs, and, despite its challenging characterization, there already exist experimental evidence   supporting its utility \cite{PhysRevB.98.085125}. Moreover, since the spatial nonlocality reflects the MBSs at spatially separated regions, it is natural to wonder if such a spatial nonlocality carries quantum correlations and influences the entanglement properties of the system \cite{PhysRevLett.120.240403}.

\begin{figure}[t]
\centering
\includegraphics[width=.8\columnwidth]{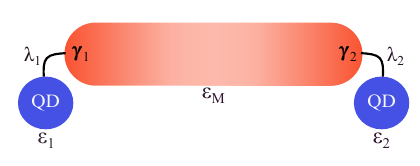}
 \caption{Sketch of the system: a topological superconductor (red) hosting MBSs $\gamma_{1,2}$ is coupled by $\lambda_{1,2}$ to two QDs (blue) with onsite energies $\varepsilon_{1,2}$. The energy associated with the spatial overlap between $\gamma_{1}$ and $\gamma_{2}$ is represented by $\varepsilon_{\rm M}$.
}
 \label{Fig1}
\end{figure}

In this work, we consider a topological superconductor with two MBSs coupled to QDs, as shown in Fig.\,\ref{Fig1}. We investigate the entanglement properties of MBSs. In particular, we characterize entanglement by exploring the dynamics of concurrence and discord, quantum correlations that we also complement by obtaining the entanglement dynamics and return probability. We find that zero-energy MBSs, which are completely localized at the edges, can evolve an initial state of maximally entangled QDs to a classical state of the system. Interestingly, we discover that a finite Majorana overlap enables maximally entangled states between MBSs and a QD, an effect that is fully controllable by the couplings to the QDs. We show that entanglement generation occurs when the initial state is maximally entangled and, surprisingly, also for an initial separable state, where maximal entanglement in the latter is achieved in the long time dynamics. In both cases, the entanglement signatures are intimately tied to energy overlap between MBSs and, therefore, associated with their spatial nonlocality.  We also show that QDs mediated by a regular fermion do not produce the entanglement signatures found for MBSs. While entanglement generation  has been explored through other mechanisms—such as due to photons \cite{PhysRevResearch.6.043029, Shadbolt_2011, Riedinger2018} and cavities \cite{PRXQuantum.5.020339, Vijayan2024}, our work exploits the spatial nonlocality of MBSs, which then   provides   an alternative way for characterizing  MBSs and highlights their potential for quantum information.  Given the current experimental advances on quantum correlations, our findings can be measured by performing quantum state tomography \cite{altepeter2004tomography,doi:10.1126/science.1130886}  which allows to reconstruct the system’s density matrix.

The paper is organised as follows: In Section II, we describe the model Hamiltonians of the systems and obtain their eigenenergies and eigenvectors. In Sec. III, we briefly describe the different measures of quantum correlations and state dynamical quantities. Following that, in Sec. IV, we compute the state dynamical functions and quantum correlation measures and give analytical expressions for all quantities for different initial states. Subsequently, we discuss the results in the later part of the section. Finally, in Section V, we summarise and conclude the results.


\section{The Model Hamiltonians and methods}
\label{Models}
We are interested in exploring quantum correlations and entanglement signatures of MBSs. For this purpose, we consider a topological superconductor hosting MBSs and couple them to two QDs, as shown in Fig.\,\ref{Fig1}. Moreover, we contrast this topological model by replacing the nonlocal fermion, arising due to both MBSs, with a regular fermion. In what follows, we present these two models, discuss their properties, and introduce the methods used in this work to quantify entanglement measures.
\subsection{QDs coupled through MBSs}
The Hamiltonian describing the two MBSs coupled to QDs is given by
\begin{equation}
\label{MajoranaH}
H = \frac{i\epsilon_M}{2}  \gamma_1\gamma_2 + \sum_{j=1,2} \epsilon_{j}d_j^\dag d_j+ \lambda_j(d_j^\dag - d_j)\gamma_j\,,
\end{equation}
where $\gamma_{i}$ represents a Majorana operator, $\epsilon_M$ is the energy of their splitting between MBSs, $d_j^\dag (d_j)$ creates (annihilates) a fermionic state in the QDs, and $\lambda_j$ characterizes the coupling between QDs and MBSs. We note that the energy splitting $\epsilon_M$ reflects the Majorana nonlocality and can be approximated  as \cite{PhysRevB.86.220506} $\epsilon_M \sim e^{-L/l}$, where L is the length of the topological superconductor and $l$ is the Majorana localisation length. Thus, $\epsilon_M = 0$ is achieved in a very long superconductor, where $L \gg l$. In realistic systems, however, a finite value of $\epsilon_M$ arises due to the overlap of the wavefunctions of the MBSs \cite{Albrecht2016}. Here, it is important to note that MBSs at zero energy represent true Majorana modes, whereas at finite energy, they can be seen to represent trivial ABSs \cite{prada2019andreev}.

To access quantum correlations and entanglement, it is convenient to write Eq.\,(\ref{MajoranaH}) in terms of fermionic operators and then handle it in number operator basis states. Thus, the two Majorana operators are transformed into a nonlocal fermion by using $\gamma_1 = i(f^\dag-f)$, $\gamma_2 = f^\dag + f$, where $f$ represents the nonlocal fermion made of Majorana operators.   Now, the two QDs plus the nonlocal fermion represent a three fermion system, with  number  operators $n_{d_i}=d^{\dagger}_{i}d_{i}$ for the QDs while $n_f=f^{\dagger}f$ for the nonlocal  fermion operator. This allows us to define a three-qubit system, which will be analyzed next.

Before proceeding, we make some simplifications by setting the on-site energies of the QDs to zero, $\epsilon_{1} = \epsilon_{2} = 0$. This choice places the QDs in resonance with the truly localized MBSs, facilitating the derivation of analytical expressions; as we will see, this regime is also crucial for generating quantum correlations in the system. Further, we set the coupling to the left QD to unity: $\lambda_1=1$, $\omega = \epsilon_M/2$, $\lambda_{2}=\lambda$, and $\hbar=1$. Hence,  tuning $\lambda$ and $\omega$ allows us to control the properties of the entire system. We now proceed to write Eq.\,(\ref{MajoranaH}) in the number representation basis defined by $\ket{n_{d_1}, n_{d_2},n_f}$. In this regard, we note that, since the Hamiltonian commutes with the fermion parity operator $(-1)^N$, with $N$ being the total number of fermions, the odd and the even sectors do not mix and  can be solved separately. We focus on  the even sector with basis states $\ket{000}$, $\ket{011}$, $\ket{101}$, and $\ket{110}$. The even Hamiltonian can be then written as,
\begin{equation}
\label{Heven}
H_{e}= \begin{pmatrix}
-\omega &  \lambda &  -1 & 0	\\
\lambda  & \omega & 0 & -1 \\
-1 & 0 & \omega &  \lambda 	\\
0 & -1  & \lambda  & -\omega 
\end{pmatrix}.
\end{equation}

The eigenvalues of the even sector are then given by
\begin{equation}
\label{EnergyMS}
E_{i}=\pm\sqrt{(\lambda\pm1)^{2}+\omega^{2}}\,,
\end{equation}
where  $E_{1(2)}=\mp\Delta$ and $E_{3(4)}=\mp\Delta_{1}$, with $\Delta = \sqrt{(\lambda-1)^2+\omega^2}$ and $\Delta_1 =\sqrt{(\lambda+1)^2+\omega^2}$. The eigenvectors $\ket{E_j}$ in increasing value of $j$ can be written as un-normalized column vectors in the order of its corresponding eigenvalues as
\begin{equation}
\label{MS}
EV = (\ket{E_1}|\ket{E_2}|\ket{E_3}|\ket{E_4})\,,
\end{equation}
where $\ket{E_1} = (1, [\omega-\Delta]/[\lambda-1], [\omega-\Delta]/[\lambda-1], 1)^T$ and T is the transpose operation. The eigenvector $\ket{E_2}=\ket{E_{1}(\Delta\rightarrow-\Delta)}$, and $\ket{E_3} = (-1, [-\omega+\Delta_1]/[\lambda+1], [\omega-\Delta_1]/[\lambda+1], 1)^T$, while $\ket{E_{4}}=\ket{E_{3}(\Delta_1\rightarrow -\Delta_{1})}$.
We note that all the eigenvalues in Eqs.\,(\ref{EnergyMS}) have contributions from both $\lambda$ and $\omega$, while the respective eigenstates in Eq.\,(\ref{MS}) exhibit contributions from all configurations the even ket states listed above Eq.\,(\ref{Heven}). This distribution of the Hamiltonian parameters in the eigenvalues and eigenvectors arises precisely from the nonlocal fermionic structure of MBSs. We note that, although the focus of the main part is on the even sector, we also discuss the odd sector in Appendix \ref{oddsector}. However, the main findings of this work are independent of the sector.

\subsection{QDs coupled through a normal fermion}
\label{NFS}
To contrast our results, here we discuss a model where the QDs are coupled via a normal fermion, which is not nonlocal and therefore of no Majorana origin.  We model this normal fermion system (NFS) by the following Hamiltonian,
\begin{equation}
\label{NormalF}
H_{\rm NFS} =  \epsilon_c c^\dag c + \sum_j \epsilon_{j}d_j^\dag d_j+ \lambda_j(d_j^\dag c + h.c.)\,,
\end{equation}
where  $\epsilon_c$ is onsite fermion energy, $c^{\dagger} (c)$ creation (annihilation) operators of the normal fermion, while $d_i^\dag (d_i)$ is the creation (annihilation) operator in the QDs. Here, the coupling between   QDs and the normal fermion is characterized by $\lambda_j$. We set $\epsilon_c = 2\omega$ so that we treat the energies of the nonlocal and regular fermion in Eqs.\,(\ref{MajoranaH}) and Eq.\,(\ref{NormalF}) at the same level. As for the Majorana system,  we solve Eq.\,(\ref{NormalF}) in the even sector spanned by number operator basis states, $\ket{n_{d_1} n_{d_2} n_c}$ and consider the same energy unit $\lambda_{1}=1$ so that the coupling to the right QD is $\lambda_{2}=\lambda$.  We then obtain the eigenvalues which for the even sector are given by
\begin{equation}
\label{EnergyNFS}
E^\prime_i= \lbrace 0, 2\omega, \Delta_{-}, \Delta_{+}\rbrace\,,
\end{equation}
where $\Delta_{\pm}=\omega\pm\sqrt{1+\lambda^{2}+\omega^{2}}$.  Moreover, the associated un-normalized eigenstates for increasing values of $i$ in $E^\prime_i$ are given by
\begin{equation}
\label{PsiNFS}
 EV_{\rm NFS} = (\ket{E_1^\prime}|\ket{E_2^\prime}|\ket{E_3^\prime}|\ket{E_4^\prime})\, ,   
\end{equation} 
where $\ket{E^\prime_1}= (1,0,0,0 )^T$, $\ket{E^\prime_2}= (0,\lambda,1,0)^T$, $\ket{E^\prime_3}= (0,1/\Delta_{+},-\lambda/\Delta_{+},1 )^T$,  and $\ket{E^\prime_4}= (0,1/\Delta_{-},-\lambda/\Delta_{-},1 )^T$. Before going further, it is worth pointing out that the first eigenvalue is zero with eigenstate  $\ket{E^\prime_1}$, which implies that it has only the configuration $\ket{000}$ where no excitations are present. In this first state, neither the onsite energy nor the hopping terms of the Hamiltonian contribute to the energy, resulting in a zero eigenvalue. The second eigenvalue solely depends on the energy associated with the normal fermion $\epsilon_{c}=2\omega$, indicating the presence of  the normal fermion in its eigenstate $\ket{E^\prime_2}$ by having $\ket{011}$ and $\ket{101}$ terms but not $\ket{000}$ and $\ket{110}$.  The last two eigenvalues depend on  the parameters of the Hamiltonian and acquire eigenstates having mixtures of all possible excitations, except $\ket{000}$. This point is of interest because  if $\ket{000}$ is not present in the initial state, it will not emerge throughout the dynamics, as we will show later.

\subsection{Methods for quantifying entanglement and state dynamics}
\label{entanglementmeasures}
We are interested in exploring the effect of nonlocality on entanglement and quantum correlations in the systems discussed in previous section. To address this task, we focus on bipartite subsystems described by  a reduced density matrix $\rho_d$ obtained as \cite{9781107002173}
\begin{equation}
\label{RDM}
\rho_d = \text{Tr}_{d^\prime}(\rho)\,,
\end{equation}
where $\rho$ is the density matrix of the composite three qubit system, and ${\rm Tr}_{d^{\prime}}$ represents the partial trace operation over a subsystem such that $\rho_d$ describes a two-qubit system. We remind that the three-qubit system is formed by the two QDs and either the nonlocal fermion due to MBSs or a normal fermion, see Section \ref{Models} for more details on the models. The density matrix in Eq.\,(\ref{RDM}) is obtained by using a standard approach 
$\rho=\ket{\psi}\bra{\psi}$, where $\ket{\psi}$ corresponds to the state of the system \cite{9781107002173}. We anticipate that, because we are interested in the dynamical properties, we use the time-evolved states of the Majorana and normal fermion systems discussed in the previous section, taking into account the different initial states in both systems. Then, all the bipartite quantum correlations can be studied by calculating the reduced density matrix $\rho_d$ given by Eq.\,(\ref{RDM}), thus providing a starting point for exploring various entanglement measures. In particular, we will address concurrence and discord because they provide a unifying way to quantify entanglement and quantum correlations \cite{RevModPhys.81.865, Bera_2018}. To complement these quantities, we will also address the state dynamics by employing the return probability and the entanglement dynamics. To make this work self contained, in what follows, we briefly describe the fundamental aspects of these quantities.

\subsubsection{Concurrence}
\label{concurrence}
The concurrence, denoted here by $C$, measures entanglement between  two qubits in a mixed state \cite{PhysRevLett.80.2245}. The concurrence is calculated as \cite{PhysRevLett.80.2245}
\begin{equation}
\label{CC}
C = \max(\xi_1-\xi_2-\xi_3-\xi_4, 0)\,,
\end{equation}
which correspond to the maximum of the eigenvalues $\xi_i$, with them being the eigenvalues in  decreasing order of the matrix $R= \sqrt{\sqrt{\rho_d} \tilde{\rho_d}\sqrt{\rho_d}}$. The matrix $\tilde{\rho_d}$ is defined as $\tilde{\rho_d}= \sigma_y\otimes\sigma_y\rho_d^{*}\sigma_y\otimes\sigma_y$, where $\rho_d^{*}$ is the complex conjugate reduced density matrix $\rho_d$ given by Eq.\,(\ref{RDM}).  In our systems, the reduced density matrices in the basis states of $\ket{00}$, $\ket{01}$, $\ket{10}$, and $\ket{11}$ acquire the following form
\begin{equation}
\label{RDMX}
\rho_d= 
\begin{pmatrix}
u & 0 & 0 & y^*	\\
0  & w_1 & x^* & 0 \\
0  & x & w_2 &  0	\\
y & 0  & 0 & v \\
\end{pmatrix}\,,
\end{equation}
which takes the form of a $X$-state because the subsystems are in even sectors, and the odd and even sectors do not mix within the system \cite{Vimal1}. We note that  the non-zero matrix elements in Eq.\,(\ref{RDMX}) represent correlation functions between the two qubits, which, as we will see in the next section, turn out to be functions of the coefficients of the eigenfunctions.  The zero elements represent the fact that the odd and even states do not mix in the system. For the reduced density matrix by Eq.\,(\ref{RDMX}), the concurrence can be written as
\begin{equation}
\label{CX}
C = 2\max(|x| - \sqrt{uv}, |y|-\sqrt{w_1w_2}, 0)\,,
\end{equation}
where the maximum of the three terms is obtained by computing the eigenvalues of the matrix $R$ given just below Eq.\,(\ref{CC}), resulting in two finite entries in Eq.\,(\ref{CX}). Before going further, a few additional comments are warranted at this stage. The concurrence given by Eq.\,(\ref{CX}) remains finite if the off-diagonal correlation functions ($x,y$) dominate over their diagonal counterparts ($\sqrt{uv}, \sqrt{w_{1}w_{2}}$); otherwise, $C$ goes to zero.  A zero concurrence $C=0$ signals the appearance of an unentangled state of the system, but, importantly, this does not imply that the total quantum correlations in the system are completely lost or absent \cite{Vimal2}. The approach discussed here will be used later to obtain concurrence and investigate entanglement in a bipartite system.

\subsubsection{Quantum discord}
\label{discord}
To quantify quantum correlations beyond  entanglement, we focus on the quantum discord because it measures  purely quantum correlations in a bipartite system.  The quantum discord is of particular relevance when exploring the dynamics of the system, which, although  acquiring a non-entangled state at certain times, may still exhibit finite quantum correlations \cite{Datta1, Lanyon1, Dakic1, Brodutch1, Pirandola1, Almeida1, Mansour1, AitMansour1}. The quantum discord is a bipartite correlation measure, and, in this sense, it is similar to concurrence. It is defined by the difference between two classically equivalent expressions of quantum mutual information \cite{Ollivier1}, which generalizes the classical mutual information to quantify the correlation between two subsystems of a bipartite system. Classically, mutual information can be defined in two equivalent ways: one using the entropies of subsystems and the other using conditional entropies \cite{Ollivier1}. In the quantum regime, the mutual information is characterized by the conditional entropy and  involves measurements on one part of the system;  the conditional entropy fundamentally differs from its classical counterpart because it depends on the chosen eigenbasis states. This creates a difference between the two expressions for mutual information in the quantum regime, which is quantified by quantum discord. It is thus useful to first introduce quantum mutual information, which, for subsystems A and B, is characterized by \cite{Ollivier1}
\begin{equation}
\label{ClassIJ}
\begin{split}
I(\rho_{AB}) &= S(\rho_A)+S(\rho_B)-S(\rho_{AB})\,,\\
J(\rho_{AB}) &= S(\rho_A)-C_{\theta,\phi}(\rho_{A|B}))\,,
\end{split}
\end{equation}
where $S(\rho)$ is the von Neumann entropy associated to $\rho$ and calculated as  $S(\rho) = -Tr(\rho\log_2\rho)$, while $C_{\theta,\phi}(\rho_{A|B})$ is the conditional entropy of A given the state of B. Moreover, $\rho_{A(B)}$ is the reduced density matrix of A(B), $\rho_{AB}$ is the composite reduced density matrix of the subsystem AB, and $\rho_{A|B}$ is the conditional density matrix of A conditioned on the measurement basis of B. Furthermore, the conditional entropy is obtained as \cite{Ollivier1}   
\begin{equation}
\label{EqC}
C_{\theta,\phi}(\rho_{A|B}) = \min_{\lbrace B_{\tilde{\kappa}}\rbrace}\sum_{\tilde{\kappa}}p_{\tilde{\kappa}}S(\rho_{A|B_{\tilde{\kappa}}})\,,
\end{equation}
where $\theta$ and $\phi$ are angles that parametrize the measurement basis of B, $\lbrace B_{\tilde{\kappa}}\rbrace = \lbrace \ket{\tilde{\kappa}}\bra{\tilde{\kappa}} \rbrace $, $\tilde{\kappa} = \lbrace \tilde{0}, \tilde{1}\rbrace$ is a complete set of projection operators corresponding to local measurements on B, and  {\rm min} indicates that a minimization operation with respect to $\lbrace B_{\tilde{\kappa}}\rbrace$ is carried out to find $C_{\theta,\phi}(\rho_{A|B})$. Note that this minimization process involves finding the values of $\theta$ and $\phi$. Also, $p_{\tilde{\kappa}}$ on the right hand side of Eq.\,(\ref{EqC}) denotes the probability of measurement outcome $\tilde{\kappa}$ and is defined as   $p_{\tilde{\kappa}}=\text{Tr}[(\sigma_0\otimes B_{\tilde{\kappa}})\rho_{AB}(\sigma_0\otimes B_{\tilde{\kappa}})]$, with $\sigma_0$ representing the identity matrix for A. Moreover,  $\rho_{\tilde{\kappa}}=(\sigma_0\otimes B_{\tilde{\kappa}})\rho_{AB}(\sigma_0\otimes B_{\tilde{\kappa}})/p_{\tilde{\kappa}}$, which represents that it is the density matrix $\rho_{AB}$ conditioned to the measurement outcomes $\tilde{\kappa}$ and weighted over different outcomes of marginal conditional entropies $S(\rho_{A|B_{\tilde{\kappa}}})$. The term $\rho_{A|B_{\tilde{\kappa}}}$ represents the conditional density matrices in the $|\tilde{\kappa}\rangle$ basis. In terms of all the above considerations, the expression for quantum discord is \cite{Ollivier1} 
\begin{equation}
\label{discordQ}
D_{AB} = {\rm min_{\lbrace B_{\tilde{\kappa}}\rbrace}}[I(\rho_{AB})- J(\rho_{AB})]\,,
\end{equation}
where $I(\rho_{AB})$ and $J(\rho_{AB})$ are given by Eq.\,(\ref{ClassIJ}). The minimization of the difference of two expressions over $\theta$ and $\phi$ in above equation defines quantum discord. Alternatively, it implies the computation of minimized conditional entropy, as defined in Eq.\,(\ref{EqC}), which makes the quantum discord a quantifier of truly quantum correlation.  In Appendix \ref{AppendixA}, we provide step-by-step calculations of discord in different subsections for both MS and NFS. Here, with above considerations, we can write the expression for quantum discord 
\begin{equation}
\label{discordexp}
D_{AB} = \min_{(\theta, \phi)} C_{\theta, \phi}(\rho_{A|B})- S(\rho_{AB})+S(\rho_B)\,,
\end{equation}
as a function of $\theta$ and $\phi$. From the above equation, it is evident that minimized conditional entropy contributes positively to quantifying discord. Therefore, a higher value of the minimized conditional entropy indicates stronger quantum discord between subsystems A and B. Additionally, the positive contribution from $S(\rho_B)$ demonstrates that stronger entanglement between the two subsystems implies greater quantum discord. However, the composite entropy $S(\rho_{AB})$ consistently contributes negatively to the discord.

\subsubsection{State dynamics}
One of the basic interests in quantum systems is to study the decay or revival of the initial state in the dynamics of the system. However, the state dynamics can also be manipulated to understand the entanglement signatures of the subsystems over time. 
Motivated by these ideas, in this work we also address the  dynamics of a maximally entangled state, called \emph{entanglement dynamics} and denoted  by $E_d$, and the return probability denoted by $R_p$. These two quantities are obtained as
\begin{equation}
\label{Statedynamics}
\begin{split}
R_p &= |\bra{\psi(0)}e^{-iHt}\ket{\psi(0)}|^2\,,\\
E_d &= |\bra{\phi}e^{-iHt}\ket{\psi(0)}|^2\,.
\end{split}
\end{equation}
As we observe, the return probability $R_{p}$ is defined  by overlapping a time-evolved state with its initial state, while the entanglement dynamics $E_d$ is defined by projecting a time-evolved state function onto a desired state $\ket{\phi}$, whose explicit form will be given later. It is worth noting that, due to the definitions in Eqs.\,(\ref{Statedynamics}), the return probability always begins with unity, regardless of the chosen initial states. However, the initial value for the entanglement dynamics depends on both the initial   and the desired states. With Eqs.\,(\ref{Statedynamics}) as well as with those for concurrence and discord, we are now in position to explore the quantum correlations and entanglement signatures for the subsystems discussed in Section \ref{Models}.


\section{Results for a maximally entangled initial state}
\label{secmaxentangled}
In this part, we follow the discussion of previous section and obtain the quantum correlations and state dynamic probabilities for the two systems described in   Section \ref{Models}, taking into account a maximally entangled initial state. We remind that the first introduced setup  corresponds to a Majorana system (MS), where QDs are coupled via nonlocal fermion of Majorana origin. The second setup consists of QDs coupled via a normal fermion, coined normal fermion system (NFS). Since the manipulation of QDs was shown to be a feasible task \cite{RevModPhys.75.1,de2010hybrid}, see also Refs.\,\cite{dvir2023realization,PhysRevX.13.031031,bordin2023crossed}, we consider that the two QDs are in a maximally entangled initial state, while the nonlocal fermion due to MBSs is in the zero state. We note that it is necessary to consider the zero state of the nonlocal fermion in order to maintain the even parity of the system, which is the sector we investigate; see Section \ref{Models}.
The separability of the non-local fermion with entangled QDs is justified by assuming that the topological superconductor and QD are not coupled before the evolution of the entire system starts.

The initial state can be thus written as
\begin{equation}
\label{initialstate1}
\ket{\psi(0)} = \frac{1}{\sqrt{2}}[\ket{000}+ \ket{110}] = \left[\frac{\ket{00}+ \ket{11}}{\sqrt{2}}\right]\ket{0}\,,
\end{equation}
which corresponds to   one of the Bell's states for the two QDs signalling that they are maximally entangled. The dynamics of the other maximally entangled states can be calculated in a similar fashion.  We anticipate that  the Bell's state $(\ket{00}-\ket{11})/\sqrt{2}$ emerges in the dynamics, while the other two  Bell's states are prohibited by the parity of the system. Then,  the state function for the initial state given in  Eq.\,(\ref{initialstate1}) is obtained as
\begin{equation}
\label{timeevolved}
\ket{\psi(t)} = e^{-iHt}\ket{\psi(0)} = \sum_{j}^4 e^{-iE_jt}\ket{E_j}\innerp{ E_j}{\psi(0)}, 
\end{equation}
where   $\ket{E_j}$   represents the eigenstates of the  MS or NFS  $(\ket{E_j}\rightarrow\ket{E_j^\prime})$ given by Eqs.\,(\ref{MS}) or Eqs.\,(\ref{PsiNFS}), depending on the system under investigation. Note that the initial state $\ket{\psi(0)}$ is given in Eq.\,(\ref{initialstate1}). Before going further,  we point out that the contribution from the eigenstates in Eq.\,(\ref{timeevolved}) also depends on the form of the initial states. In this regard, we have that the primary contributions in the MS come from eigenstates $\ket{E_1}$ and $\ket{E_2}$ with energies $-\Delta$ and $\Delta$, respectively. The eigenstates $\ket{E_3}$ and $\ket{E_4}$ are orthogonal to the initial state, $\innerp{ E_j}{\psi(0)}=0$, which implies that they do not contribute to the dynamics given by Eq.\,(\ref{timeevolved}).  Contrary to the MS, the dynamics of the NFS is determined by the contributions of all states.

\subsection{Majorana system}
\label{MajoSysMax}
In the case of the system with MBSs, the time-evolved state given by Eq.\,(\ref{timeevolved}) acquires the  form given by
 \begin{equation}
 \label{psit}
 \begin{split}
\ket{\psi(t)} &= \frac{1}{\sqrt{2}}
\ket{0}( \eta\ket{00} + \chi\ket{11})\\
&+
\frac{1}{\sqrt{2}}\ket{1}(\eta\ket{10} + \chi\ket{01})\,,
\end{split}
\end{equation}
where $\chi$ and $\eta$ are time-dependent and given by
\begin{equation}
\label{etachi}
\begin{split}
\eta(t)&=\frac{1}{\Delta}(\Delta\cos{\Delta t} + i \omega\sin{\Delta t})\,,\\
\chi(t)&= \frac{-i}{\Delta}(\lambda-1)\sin{\Delta t}\,.
\end{split}
\end{equation}
Here, $\Delta$, $\lambda$, and $\omega$ represent the lowest positive eigenvalue, the coupling between nonlocal fermions and the right QD, and the energy splitting between MBSs, as described in Subsection \ref{MS}. The coefficients $\eta$ and $\chi$ in the above state function shape the dynamics of the entanglement generations between the MBSs and the second QD. They can also be manipulated to see the type of maximally entangled state (of the Bell's state form) that may be achieved in the evolution. Therefore, below, we now compute the quantities discussed in Section \ref{entanglementmeasures}.

\subsubsection{Return probability and entanglement dynamics}
By using Eq.\,(\ref{Statedynamics}), we obtain the entanglement dynamics $E_d$ and the return probability $R_{p}$. For  $E_d$  we need to specify the desired state onto which will be projected the time-evolved function $\ket{\psi(t)}$. For this purpose, we consider $\ket{\phi}$  describing a maximally entangled  nonlocal fermion and the second QD given by
\begin{equation}
\label{phi}
 \ket{\phi}= \frac{1}{\sqrt{2}}[\ket{000}+\ket{011}]= \ket{0}\frac{1}{\sqrt{2}}(\ket{00}+\ket{11})\,,  
\end{equation}
where  the second equality  expresses the essence of the considered maximal entanglement. Then, using Eqs.\,(\ref{Statedynamics}), along with Eq.\,(\ref{timeevolved}), Eq.\,(\ref{initialstate1}), and Eq.\,(\ref{phi}), we find that the return probability and entanglement dynamics are given by
\begin{equation}
\label{MSResults}
\begin{split}
 R_p &= |\eta|^2\,,\\
 E_d &= \Big|\frac{\eta+\chi}{2}\Big|^2\,,
\end{split}
\end{equation}
where $\eta$ and $\chi$ correspond to the time-dependent coefficients of the time-evolved state and given by Eqs.\,(\ref{etachi}). Interestingly, we see that the entanglement dynamics $E_{d}$ is sensitive to the local phases in the state function, unlike the return probability.  This sensitivity is crucial for visualizing other maximally entangled states generated during the evolution. 

\subsubsection{Concurrence}
We are interested in computing the concurrence between the two QDs for the initial state written in Eq.\,(\ref{initialstate1}). For this purpose, we follow the recipe discussed in Subsection \ref{concurrence}, which involves  obtaining the density matrix elements associated to the time-evolved state Eq.\,(\ref{psit}) and then using Eq.\,(\ref{CX}). We obtain the density matrix and find that  its elements are given by $u = v = y =|\eta|^2$/2, $ w_1= w_2 = x = |\chi|^2/2$. Therefore, using Eq.\,(\ref{CX}), we find that the concurrence is given by 
\begin{equation}
\label{CMS1}
C = ||\eta|^2-|\chi|^2|\,,
\end{equation}
where the modulus in the expression on the right-hand side arises from the fact that particular sets of diagonal and off-diagonal elements of the density matrix have the same expression.  From Eq.\,(\ref{CMS1}), we identify that the concurrence vanishes with $C=0$ only when $|\eta|=|\chi|$. It is worth noting that   zero concurrence has many properties, including the signature of entanglement monogamy \cite{Terhal2}, which states that the entanglement cannot be freely shared between more than two parties. We will provide a deeper analysis of the zero concurrence regime  by
  comparing it with the entanglement dynamics and the return probability later in this section.

\subsubsection{Quantum discord}
To compute quantum discord in MS, we follow the discussion provided in Subsection \ref{discord} along with Eq.\,(\ref{discordQ}) to obtain the discord between the two QDs represented by A and B. Detailed calculations are carried out in the first subsection of Appendix \ref{AppendixA}, formulating the quantum discord as follows: 
\begin{equation}
D_{AB} = \min_{(\theta, \phi)} C_{\theta, \phi}(\rho_{A|B})- S(\rho_{AB})+S(\rho_B)\,,
\end{equation}
where the term $C_{\theta, \phi}(\rho_{A|B})$ represents the conditional entropy of the first QD when the measurement basis of the second QD is parameterized by angles $\theta$ and $\phi$. The term $S_{AB}$ denotes the composite entropy of the two QDs, while $S_B$ is the single-qubit entropy of the second QD. Further details of the calculation is given in Appendix \ref{AppendixA}.

\subsection{Normal fermion system}
\label{NFSMaxEnt}
In this part, we carry out the same calculations as in the previous subsection but when the QDs are coupled via a normal fermion.  The time-evolved state function for the initial state from Eq.\,(\ref{initialstate1}) is here given by
\begin{equation}
\label{NFSPSI}
\ket{\psi(t)} = \bar{c}_1\ket{000}+\bar{c}_2\ket{011}+\bar{c}_3\ket{101}+\bar{c}_4\ket{110},
\end{equation}
where $\bar{c}_{i}$ are time-dependent  coefficients   given by 
\begin{equation}
\label{coefNFS}
\begin{split}
\bar{c}_1(t)&= \frac{1}{\sqrt{2}}\,,\\
\bar{c}_2(t)&= \frac{1}{2\sqrt{2(1+\lambda ^2+ \omega^2)}}(e^{-i\Delta_{-} t}-e^{-i\Delta_{+} t})\,, \\
\bar{c}_3(t)&= \frac{-\lambda}{2\sqrt{2(1+\lambda ^2+ \omega^2)}}(e^{-i\Delta_{-} t}-e^{-i\Delta_{+} t})\,,\\
\bar{c}_4(t)&= \frac{1}{{2\sqrt{2(1+\lambda^2+\omega^2)}}}(\Delta_{+}e^{-i\Delta_{-} t}-\Delta_{-} e^{-i\Delta_{+} t})\,.
\end{split}
\end{equation}
Hence, having all the coefficients finite implies that $\ket{\psi(t)} $ for the NFS has a general and complex structure, which makes it difficult to simplify the expressions of the entanglement measures in the system, unlike what we found for the MS in the subsection \ref{MajoSysMax}. Then, we find that the return probability, entanglement dynamics, and concurrence are given by
\begin{equation}
\label{NFSmaxintRpEdCD}
\begin{split}
    R_p &= \frac{|\bar{c}_1(t)+\bar{c}_4(t)|^2}{2}\,,\\
    E_d &= \frac{|\bar{c}_1(t)+\bar{c}_2(t)|^2}{2}\,,\\
    C&= 2(||\bar{c}_1(t)||\bar{c}_4(t)|-|\bar{c}_2(t)||\bar{c}_3(t)||)\,,
\end{split}
\end{equation}
while the discord acquires a more complicated form, whose details are discussed in Appendix\,\ref{AppendixA2}. We note that to calculate $E_{d}$, we consider Eq.\,(\ref{phi}) as the desired state onto which the time-evolved state Eq.\,(\ref{NFSPSI}) was projected. For the NFS, Eq.\,(\ref{phi}) has a slightly different meaning because  the last qubit represents the normal fermion instead of the nonlocal fermion due to MBSs. Moreover, to find $R_{p}$, the initial state is taken as in Eq.\,(\ref{initialstate1}). These considerations facilitate the comparison of NFS with MS. Even though the expressions in Eq.\,(\ref{NFSmaxintRpEdCD}) acquire a complex form since $\bar{c}_{i}$ are not simple at all, we will point out their simplification  in special cases during the discussion of our results.

\begin{figure*}
\centering
\includegraphics[width=0.96\textwidth, height=0.46\textwidth]{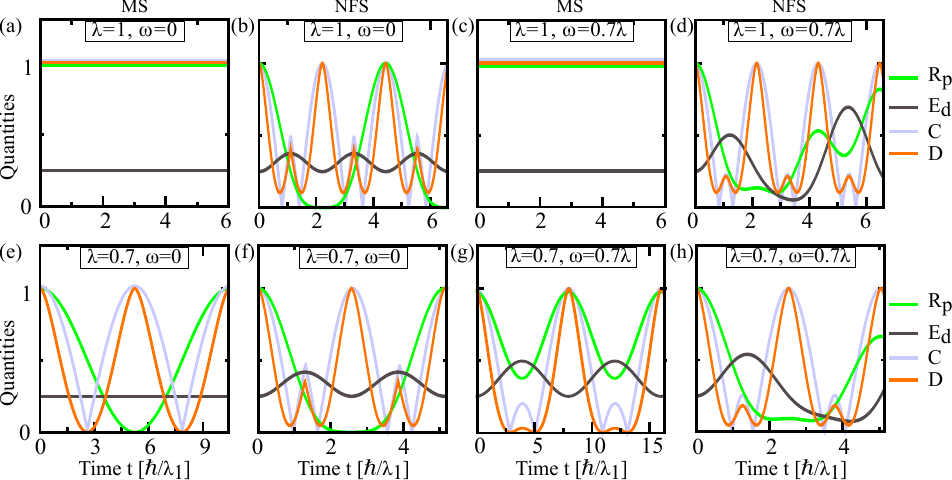}
\caption{Time evolution of concurrence, discord, return probability, and entanglement dynamics, denoted by $C$, $D$, $R_{p}$, and $E_{d}$, with  
an initial state  of maximally entangled QDs for   MS and NFS. Top row corresponds to a symmetric coupling to the left and right QDs $\lambda=1$, with MBSs and normal fermion having zero energy $\omega =0$ in (a,b), while MBSs and normal fermion having finite energy $\omega \neq 0$ in (c,d). Bottom row shows the same as in top row but for left and right QDs having asymmetric couplings to MBSs and normal fermion, namely,  $\lambda \neq 1$. }
\label{Fig2}
\end{figure*}

\subsection{Discussion of state dynamics and quantum correlations}
Having presented the general expressions obtained of concurrence (C), discord (D), entanglement dynamics ($E_{d}$) and return probability  ($R_{p}$)   for the MS and the NFS with  maximally entangled QDs, here we discuss their time evolution.   To vizualize the time evolution, in  Fig.\,\ref{Fig2} we plot C, D, $E_{d}$, and $R_{p}$ as a function of time $t$ for the two systems under consideration (MS and NFS).  Fig.\,\ref{Fig2}(a-d) and Fig.\,\ref{Fig2}(e-h) correspond to QDs 
with symmetric ($\lambda=1$) and  asymmetric ($\lambda\neq1$) couplings 
 in the MS or NFS. As discussed before, in the MS setup, QDs are coupled via MBSs, while QDs are coupled via a normal fermion in the NFS. To truly exploit the inherent Majorana nonlocality, in Fig.\,\ref{Fig2} we contrast regimes at zero and finite energies.

We first analyze the case of symmetric couplings at zero energy, which corresponds to $\omega=0$ and $\lambda=1$ in Fig.\,\ref{Fig2}(a,b). In the case of the MS,  this regime has zero-energy MBSs and the first feature to notice is that the quantities $C$, $D$, and $R_{p}$ remain at $1$, while  $E_{d}=0.25$, as shown in Fig.\,\ref{Fig2}(a), see also subsection \ref{MajoSysMax}. Since the initial state involves maximally entangled QDs, having a constant   time evolution reveals that such a maximal entanglement is not affected.
The constant time evolution can be understood by noting that the eigenstates of the Hamiltonian given by Eq.\,(\ref{MS}), except for the third eigenstate, are orthogonal to the initial state given by Eq.\,(\ref{timeevolved}), giving a vanishing overlap that does not contribute to the dynamics as seen in Fig.\,\ref{Fig2}(a). The third eigenstate of the MS is the same as the initial state [Eq.\,(\ref{initialstate1})] and has zero  energy [Eq.\,(\ref{EnergyMS})], implying that neither the quantities nor the state evolve at all. 

In contrast to the MS setup, the quantum correlations and state dynamics for the NFS case exhibit an oscillatory behaviour with time, which starts from unity, reflecting the initial maximally entangled QDs (see Fig.\,\ref{Fig2}(b)).  The distinct time evolution occurs because the contributing eigenvalues in Eq.\,(\ref{PsiNFS}) are finite for the NFS, implying that all the coefficients from Eqs.\,(\ref{coefNFS}) contribute to the time evolution that becomes oscillatory over time.  As time progresses, we observe that there are times where $C=0$ but $D\neq0$,  signalling that  there is a finite quantum correlation between the two QDs  even when the NFS reaches its separable state, whose specific form is not evident. Also,  Fig.\,\ref{Fig2}(b)  shows that there are other times $t$ at which $C=1$ and $D=1$ when $R_{p}=0$, a regime showing that additional maximally entangled states of the QDs are being generated during the dynamics in the NFS \footnote{At $t = (\pi/\sqrt{2})\,\hbar/\lambda_1$, the coefficients $\bar{c}_2(t)$ and $\bar{c}_3(t)$ are zero and $\bar{c}_4(t) = -1$; therefore, the state of the system becomes $(\ket{000}-\ket{110})/2$, which is also a maximally entangled state of the QDs, thus giving $C=1$.}. For the return probability and entanglement dynamics of the NFS, their oscillatory behaviour can be easily seen analytically from Eq.\,(\ref{NFSmaxintRpEdCD}): $R_p= |(1+ \cos{\sqrt{2}t})/2|^2$ with a period $T= \sqrt{2}\pi\,\hbar/\lambda_1$, where $\hbar/\lambda_1$ is the unit of time. Similarly, the entanglement dynamics is given by $E_d= |(1+ i\sqrt{2}\sin{\sqrt{2}t})/2|^2$ with period $T=(\pi/\sqrt{2})\,\hbar/\lambda_1$. It is straightforward to get $R_p=1$ and $E_d= 1/4$ at $t=0$, while at  $t = ({n\pi}/{\sqrt{2}})\, \hbar/\lambda_1$ we get $R_p=0$ and  $E_d$ develop minima, leading to a fall of the state into the eigenstate $\ket{000}$, as depicted by the green and black curves in Fig.\,\ref{Fig2}(b). The  overall behaviour of the entanglement measures in the NFS setup is clearly different to what we found for the MS. Therefore, zero-energy MBSs and zero-energy normal fermion induce distinct entanglement signatures.
 
In the case of finite energies $\omega \neq 0$ at symmetric couplings $\lambda=1$, the time evolution of the entanglement measures exhibits a rather similar behaviour as the one for zero energy discussed in the previous two paragraphs; see Fig.\,\ref{Fig2}(c,d). For the MS in Fig.\,\ref{Fig2}(c), we find $R_{p}$, $C$, and $D$  exhibiting a constant value equal to $1$, while $E_{d}=0.25$, which is the same as what we saw in Fig.\,\ref{Fig2}(a) at $\omega=0$. 
To understand the intriguing behaviour at $\omega \neq 0$, we note that  the coefficient $\chi$ [Eq.\,(\ref{etachi})] of the time evolved state given by Eq.\,(\ref{psit}) vanishes at $\lambda=1$ while the other coefficient becomes $\eta(t)={\rm cos}(|\omega|t)+i{\rm sin}(|\omega|t)$ and is thus entirely determined by the energy splitting of MBSs. As a result, the time-evolved state from Eq.\,(\ref{psit}) evolves with a constant phase, determined by $\omega$, where $|\eta(t)|^{2}=1$ enables $C$, $D$, and $R_{p}$ to remain constant at $1$ and $E_{d}=0.25$, see Eq.\,(\ref{etachi}). The quantities equal to $1$ then imply that the system remains with maximally entangled QDs and $E_{d}=0.25$ that  there is only a contribution from the configuration $\ket{000}$ in the entanglement dynamics. For the NFS at $\lambda=1$ and $\omega \neq 0$ in Fig.\,\ref{Fig2}(d), the entanglement measures exhibit a time evolution that is similar to the $\omega = 0$ case but with some subtle  differences. At $\omega\neq0$, the periodic $C$ and $D$ develop maxima, which, however,  are not accompanied by $R_{p}=0$, implying that there is no revival of the initial state at short times \footnote{We have verified that the revival of the initial state occurs over a longer period of time than shown in the plot.}.  This situation implies  that another entangled state, having a component of the initial state, is being   generated during the evolution.  Moreover, the  minima of $E_d$ states that one of the configurations in $\ket{\phi}$ remains in the state function throughout the dynamics. 

When the couplings become asymmetric, $\lambda \neq 1$, the entanglement measures of the MS as well as those of the NFS develop an oscillatory profile; see Fig.\,\ref{Fig2}(e-h). At $\omega=0$, the MS with zero-energy MBSs achieves $D=0$ and $C=0$ at certain times, which is contrary to the NFS case, where $D\neq0$ when $C=0$, see Fig.\,\ref{Fig2}(e,f). The evolution of the MS in Fig.\,\ref{Fig2}(e) starts from the state $(|000\rangle + |110\rangle)/\sqrt{2}$, which has maximally entangled QDs maintaining the even parity of the entire system. Since the system is initially in a maximally entangled state between the two QDs, it transitions from maximal entanglement by reducing the contribution of the configuration $\ket{110}$ and moves towards the separable state $\ket{000}$. The system's state is well-characterized as the entanglement dynamics appropriately yield $E_d = 1/4$, due to the contribution solely from the configuration $\ket{000}$. As the evolution progresses beyond this point, the configuration $\ket{110}$ acquires a finite phase that  eventually becomes $-1$; this process then results in the state $(|000\rangle - |110\rangle)/\sqrt{2}$, which is again a state with maximally entangled QDs characterized by $C=1$, $D=1$, $R_p=0$, and $E_d= 1/4$.

It is important to note that the dynamics beyond the point when the system reaches $\ket{000}$ involves only the additional configuration $\ket{110}$, as   other possibilities  such as $\ket{011}$ and $\ket{101}$, are excluded. In fact, the configuration $\ket{011}$ does not contribute because $E_d$ remains constant, and $\ket{101}$ does not contribute because, physically, this would imply the generation of entanglement between the first QD and the nonlocal fermion from the MBSs, which would decrease the entanglement between the two QDs due to the monogamy property of entanglement. However, since the dynamics indicates increasing entanglement beyond the time when the system reaches $\ket{000}$, we can be sure that only $\ket{110}$ is present in the dynamical state of the system.

Contrasting the constant value of $E_d=0.25$, it oscillates for the NFS, giving distinct signatures of both systems [Fig.\,\ref{Fig2}(e,f)]. At finite energies ($\omega\neq0$) and asymmetric couplings ($\lambda \neq 1$), the MS and the NFS exhibit similar properties as seen in  Fig.\,\ref{Fig2}(g,h) but still with some small differences. Similar behavior is associated with the oscillatory profile, which stems from having extra configurations participate in the dynamics because the coefficient of the time-evolved state in Eq.\,(\ref{etachi}) is non zero, $ \chi \neq 0$.  It is worth noting that  the entanglement measures for the MS setup have larger periodicity [Fig.\,\ref{Fig2}(g)], which occurs  because the eigenenergies in Eq.\,(\ref{MS}) have contributions from $\lambda$ and $\omega$; see also Eqs.\,(\ref{etachi}) and Eq.\,(\ref{psit}). Among the differences at $\omega\neq0$ and $\lambda \neq 1$, we find that the Majorana system MS achieves regimes with $C$, $D$, and $R_{p}$ equal to $1$ (as shown in Fig.\,\ref{Fig2}(g)), which means that the finite $\omega$ prefers the revival of the initial state rather than changing the local phase, which would result in achieving another maximally entangled state of quantum dots, as depicted for the MS in Fig.\,\ref{Fig2}(e). This phenomenon, however, is not observed in the NFS, as shown in Fig.\,\ref{Fig2}(h). Another difference is that in the NFS case, the quantities $E_d$ and $R_p$ exhibit a fall that reveals the presence of arbitrary configurations in the state; their peaks show that the different entangled states between the two QDs are being generated and destroyed. Thus, even though the oscillatory profiles of the entanglement measures for the MS and NFS exhibit some similarities, there are still significant differences that could allow identifying the MBSs.

The characteristics of MS in $\omega \neq 0$ provide further insight into the distinguishing signatures of MBSs and ABSs. It is important to emphasize that MBSs form a nonlocal fermion that is sensitive to changes in the length of the superconductor. In contrast, ABSs often have finite energies and do not depend on the length of the system; ABSs are not nonlocal. As a result, the finite energy splitting $\omega$ mimics the behavior of ABS energy. As shown in Fig.\,\ref{Fig2}, the correlation dynamics at $\omega \neq 0$ and asymmetric couplings exhibit an oscillatory time evolution that is similar for both MS and NFS, see Fig.\,\ref{Fig2}(g,h). This supports the idea that ABSs, due to $\omega \neq 0$, are local states, much like a normal fermion. Thus, the consideration of $\omega = 0$ and $\omega \neq 0$ highlights the distinct features of true zero-energy MBSs and ABSs, respectively.

Before closing this part, we highlight that the periodic characteristic of the state dynamics suggests the possible generation of maximally entangled states. Since we consider an initial state of maximally entangled QDs, it is natural to ask whether it is possible to generate maximal entanglement between other parts of the system, such as between MBSs and QDs. We address this  question  in the next subsection.

\subsection{Generating a maximally entangled state between MBSs and a QD}
As pointed out before, the oscillatory behaviour of the entanglement measures as functions of time implies a possible generation of maximally entangled states. Since the considered systems  are composed of three subsystems (qubits), with an initial state of maximally entangled QDs, we focus on   achieving  a maximally entangled state between MBSs and the right QD. To find a maximally entangled state, it is necessary to have  $\eta=\chi$ in Eq.\,(\ref{etachi}), because  this conditions places the time-evolved state given by Eq.\,(\ref{psit})  in the form of two qubit Bell's states, which are maximally entangled states \cite{9781107002173}. Thus, taking this condition into account, we find the required parameters for achieving maximal entanglement. Then, we can   choose $t = ((2n+1)\pi/{2\Delta})\, \hbar/\lambda_1$ so that $\eta$ is purely imaginary; then, by comparing  the imaginary parts of $\chi$ and $\eta$, we find that they are equal when $\omega = 1-\lambda$, which, using the expression for $\Delta$ below Eq.\,(\ref{EnergyMS}), gives $\Delta = \sqrt{2}|1- \lambda|$. Under these conditions, we obtain  $\eta=\chi={1}/{\sqrt{2}}$, which is expected to lead to a maximally entangled state between MBSs and QDs because, as we explained above, the time-evolved state acquires the form of two-qubit Bell states. 

 \begin{figure}[!ht]
\centering
\includegraphics[width=0.48\textwidth, height=0.46\textwidth]{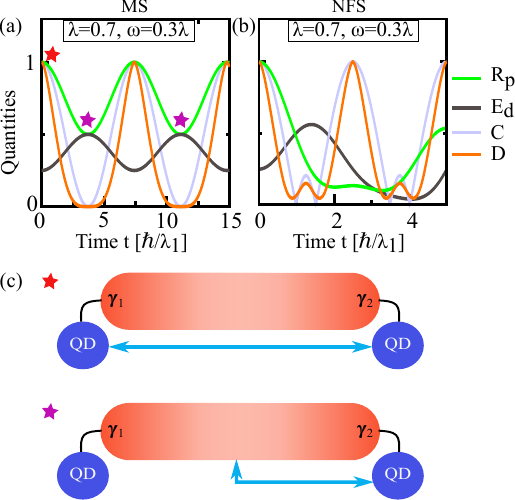}
\caption{Generation of a maximally entangled state between MBSs with a QD.
(a,b) Time evolution of concurrence, discord, return probability, and entanglement dynamics, denoted by $C$, $D$, $R_{p}$, and $E_{d}$, with  
an initial state  of maximally entangled QDs for   MS and NFS. Here, $\lambda = 0.7$ and $\omega = 1 - \lambda$ in  panels (a,b). The red star in (a) shows that the QDs are maximally entangled at the beginning of the evolution $t=0$, whereas the magenta stars show that maximal entanglement between MBSs and the second QD is achieved at later times. (c) Schematic  representation of the  entangled systems indicated by red and magenta stars in (a), with the thick cyan arrow showing the entangled parts. The NFS does not have this characteristic.}
\label{Fig3}
\end{figure}

In spite of the seemingly stringent conditions to achieve maximal entanglement during the evolution, in Fig.\,\ref{Fig3} we show that it is possible at  $\omega=1-\lambda$ and $\lambda=0.7$ by plotting the time evolution of $R_{p}$, $E_{d}$, $C$, and $D$ for the MS and NFS setups.   Apart from the already seen oscillatory behaviour, there are two points we would like to highlight, which are marked by red and magenta stars in Fig.\,\ref{Fig3}(a) for the MS setup. First, at the beginning of the evolution $t=0$, we have $R_{p}=1$, $C=1$, $D=1$, and $E_{d}=0.25$, a regime that corresponds to the maximally entangled QDs. Second,  at times given by $t = \pi/{2\Delta} = 3.7\,\hbar/\lambda_1 $, the entanglement dynamics and return probability become $E_{d}=0.5$ and $R_{p}=0.5$,   implying the creation of a maximally entangled state of MBSs and the second QD, see   magenta star in Fig.\,\ref{Fig3}(a). This maximal entanglement is further confirmed by getting  vanishing concurrence $C=0$ at such times, a situation that displays the monogamy behaviour of entanglement: when MBSs and the right QD are maximally entangled, the right QD cannot be entangled with the left QD, thus leading to $C=0$. Interestingly, the vanishing concurrence of the MS ($C=0$) is accompanied by vanishing discord ($D=0$), which reveals the absence of quantum correlations and that the system transitions into a classical state, an effect only obtained in the MS but not in the NFS; see Fig.\,\ref{Fig3}(a,b).

The key difference in achieving the classical state in the MS, as opposed to the NFS, originates from the energy contribution arising from $\omega\neq0$. This drives the dynamics in a special way that the time-evolved state can be expressed in the Bell's states form with the same coefficients for odd and even parity of the first QD, see Eq.\,(\ref{psit}). In contrast, the energy contribution in the NFS due to the normal fermion creates an arbitrarily evolving state, where each configuration has different amplitudes, as seen in Eq.\,(\ref{NFSPSI}). These distinct time-evolved states for the two systems lead to differences in the evolution of quantum correlations over time. In the MS, the classical state is achieved at different times when the conditional entropy, which captures nonclassical correlations, goes to zero, coinciding with the vanishing of entanglement. Meanwhile, in the NFS, the conditional entropy remains significant at those same moments, quantifying the nonclassical correlations in the system.

To further visualize the achieved entanglement between MBSs and right QD, in Fig.\,\ref{Fig3}(c) we schematically illustrate  the entanglement shift that corresponds to the two times of the evolution indicated by red and magenta stars in Fig.\,\ref{Fig3}(a). This entanglement shift reveals the possibility of starting with two maximally entangled QDs and then generating maximal entanglement between MBSs and the right QD during the time evolution.
In contrast to the MS system, the entanglement measures for the NFS exhibit a profile where the generation of maximal entanglement between the local fermion and the QDs cannot be determined, as shown in Fig. 3(b). Their evolution closely resembles that depicted in Fig. 2(b, d, f, h). Thus, we conclude that maximal entanglement between MBSs and QDs can be induced only in the MS. However, this requires finite-energy MBSs, which are not fully nonlocal and therefore may not be ideal for encoding information nonlocally.


\begin{figure*}
\includegraphics[width=0.96\textwidth, height=0.46\textwidth]{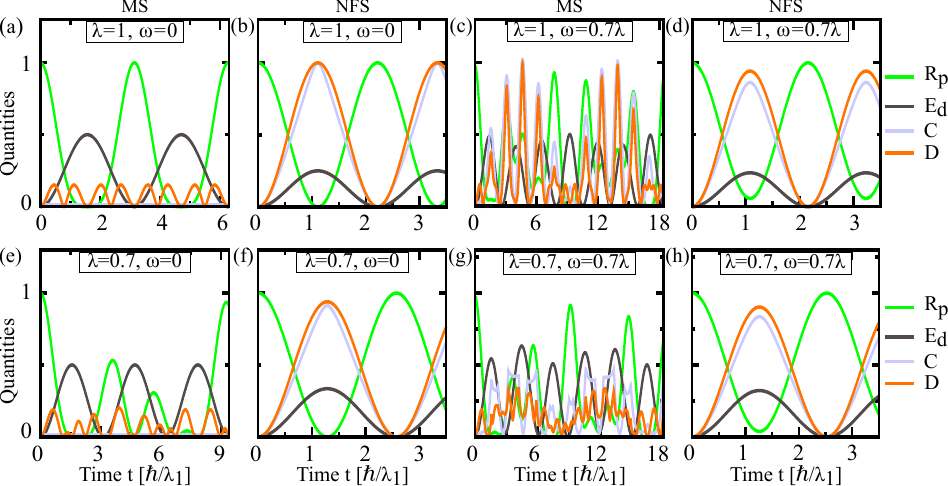}
\caption{Time evolution of concurrence, discord, return probability, and entanglement dynamics, denoted by $C$, $D$, $R_{p}$, and $E_{d}$, with  
a separable initial state for   MS and NFS.    (a, b) at $\lambda=1$ and $\omega =0$; (c, d) at $\lambda=1$ and $\omega \neq 0$; (e, f) at $\lambda \neq 1$ and $\omega =0$; and (g, h) at $\lambda \neq 1$ and $\omega \neq 0$.} 
\label{Fig4}
\end{figure*}

\section{Results for a separable initial state}
\label{secseprableinitialstate}
Having studied the state dynamics and quantum correlations for a maximally entangled initial state, in this section we focus on the same quantities but taking an initial state that is completely separable in both systems,  the MS and NFS. Thus, using the same notation as before, we consider the initial state to be given by
\begin{equation}
\label{separablest}
\ket{\psi(0)}= \ket{110}\,,
\end{equation}
which represents a separable state because it can be written as $\ket{110}=\ket{1}\ket{1}\ket{0}$. This state has excitations in the two QDs, but  the nonlocal fermion (or normal fermion when dealing with the NFS) is in the $\ket{0}$ state. We note that this separable state is also realistic because the two QDs can be initially prepared to be occupied, with a vanishing coupling to the topological superconductor controlled by voltage gates \cite{dvir2023realization,PhysRevX.13.031031,bordin2023crossed}, see also Refs.\,\cite{RevModPhys.75.1,de2010hybrid}. Moreover, we consider the $\ket {0}$ state of the nonlocal fermion because it preserves the even parity of the system. To maintain consistency with the initial state of Section \ref{secmaxentangled} and our findings therein, we do not consider the other two separable states, $\ket{011}$ and $\ket{101}$, because they require the nonlocal fermion (or local fermion in the NFS) in the $\ket{1}$ state. Furthermore, we do not address state $\ket{000}$ because this state becomes an eigenstate for the NFS, implying that the state will evolve with a constant phase which does not affect the entanglement measures over time. The desired state for obtaining the entanglement dynamics using Eq.\,(\ref{Statedynamics}) is also taken to be the same as in Eq.\,(\ref{phi}). For the separable initial state, the expressions of the entanglement measures for the MS and NFS cannot be expressed in a simplified form. Nevertheless, we list the time-dependent coefficients of the time evolved states that are required for obtaining quantum correlations and state dynamics, as discussed in Section \ref{secmaxentangled}.

 We calculate the time evolution of the $\ket{110}$ state for the MS and NFS systems, which can be written in a similar form as $\ket{\psi(t)}$ in Eq.\,(\ref{NFSPSI}). The coefficients of evolution for the MS are given by
\begin{equation}
\label{cofMSsep}
\begin{split}
c_{1(4)}(t)&=
\frac{1}{2}
  \Big\{
  \Big[\cos{\Delta t}+i\frac{\omega}{\Delta}\sin{\Delta t}\Big]\\
  &\mp\Big[\cos{\Delta_1 t}+i\frac{\omega}{\Delta_1}\sin{\Delta_1 t}\Big]
  \Big\}\,, \\
c_{2(3)}(t)&=
\frac{i}{2}
\Big\{
\frac{-(\lambda-1)}{\Delta}\sin{\Delta t}\\
&\pm\frac{(\lambda+1)}{\Delta_1}\sin{\Delta_1 t}
\Big\}\,, \\
\end{split}
\end{equation}
while for the NFS we obtain
\begin{equation}
\label{cofNFSsep}
\begin{split}
\bar{c}^\prime_1(t)&= 0\,,\\
\bar{c}^\prime_2(t)&= \frac{1}{2\sqrt{1+\lambda ^2+ \omega^2}}(e^{-i\Delta_{-} t}-e^{-i\Delta_{+} t})\,, \\
\bar{c}^\prime_3(t)&= \frac{-\lambda}{2\sqrt{1+\lambda ^2+ \omega^2}}(e^{-i\Delta_{-} t}-e^{-i\Delta_{+} t})\,,\\
\bar{c}^\prime_4(t)&= \frac{1}{{2\sqrt{1+\lambda^2+\omega^2}}}(\Delta_{+}e^{-i\Delta_{-} t}-\Delta_{-} e^{-i\Delta_{+} t})\,,
\end{split}
\end{equation}
where $\lambda$ characterizes the coupling to the right QD, $\omega$ is the energy of MBSs (normal fermion), $\Delta$, $\Delta_{1}$, and $\Delta_{\pm}$ are given below Eqs.\,(\ref{EnergyMS}) and Eqs.\,(\ref{EnergyNFS}). It is important to note that all the coefficients for MS setup are finite, and hence a complex time evolution is expected. Moreover, contrary to the coefficients of $\ket{\psi(t)}$ for initially maximally entangled states of the QDs in Eq.\,(\ref{coefNFS}) for NFS, the first coefficient, $\bar{c}^\prime_1(t)$, goes to zero. This occurs because the state $\ket{000}$ is an eigenstate of the system, and the initial state does not contain the configuration $\ket{000}$. Therefore, it does not contribute to the evolution (see Eq.\,(\ref{timeevolved})). The other coefficients are non-zero and show a similar form of dependence on $\lambda$ and $\omega$, as we will observe in the dynamics. Using the coefficients from Eq.\,(\ref{cofMSsep}) and Eq.\,(\ref{cofNFSsep}) for MS and NFS, and following the steps to calculate the state dynamical function and quantum correlations given in Section \ref{secmaxentangled}, we obtain $E_d$, $C$, and $D$ and discuss their time evolution next.

\subsection{Discussion of state dynamics and quantum correlations}
After obtaining the entanglement measures for a separable initial state, we plot them in Fig.\,\ref{Fig4} as a function of time for symmetric and asymmetric couplings between QDs and MBSs (normal fermion) in the MS (NFS) setup. To inspect the nonlocal nature of MBSs, we consider $\omega=0$ and also $\omega\neq0$ in the two cases and for the two systems. 

In the case of symmetric couplings ($\lambda=1$),  the immediate   observation is that almost all the quantities oscillate with time, with different patterns at $\omega=0$ and $\omega\neq0$, which is different to what we observed in the previous section for initial maximally entangled QDs. For the MS case in Fig.\,\ref{Fig4}(a), the quantities $R_{p}$, $E_{d}$, and $D$ oscillate, while $C=0$, implying that the completely nonlocal zero-energy MBSs cannot generate entanglement in this case. For the entanglement dynamics and return probability, we obtain $E_d = \sin^2 (t)/2$ and $R_p= \cos^2t$, which, at $t=(\pi/2)\,\hbar/\lambda_1$, become $E_d = 0.5$ and $R_p=0$, as depicted by black and green curves in Fig.\,\ref{Fig4}(a). By obtaining  $c_1(t) =1$, we conclude that at those points in time, the state $\ket{000}$ is present.  In relation to the discord $D$ in Fig.\,\ref{Fig4}(a), it exhibits homogeneous oscillations as a function of time, with period equal to $T=(\pi/4)\, \hbar/\lambda_1$, acquiring vanishing values at points where $C$ vanishes, a phenomenon we have seen to occur only for MS in Figs.\,\ref{Fig2} and Fig.\,\ref{Fig3} of the previous section. The times at which the discord completes a period ($\pi/4$) are special because the eigenvalues of the marginal entropies  are $\lbrace 0, 1\rbrace$ at $\theta=\pi/4$ and $\phi=\pi/4$, which then gives vanishing conditional entropy when obtaining the discord in  Eq.\,\ref{EqC}, see also  Eq.\,(\ref{margiEnt}) and     Eq.\,(\ref{discoQDs}) in Appendix \ref{AppendixA1}. Moreover, the composite entropy and the single QD entropy are the same and equal to unity, thus the total algebraic sum goes to zero; see details in Eq.\,(\ref{discoQDs}) in Appendix\,\ref{AppendixA1}. When analyzing the time $t=(\pi/2)\, \hbar/\lambda_1$, we have  $c_1(t)=1$ in Eq.\,(\ref{cofMSsep}) which gives a pure state description, where all entropies go to zero.  For a finite overlap ($\lambda=1, \omega=0.7$), plotted in Fig.\,\ref{Fig4}(c), an extra frequency in oscillation is introduced because of $\omega$. The entanglement is generated between QDs with fast and slow frequencies. In this case, $C$ and $D$ reach unity simultaneously, implying that maximum quantum correlation can be created by MBSs at finite energy splitting \cite{Li1}. We conclude that, while quantum correlations can be generated with zero-energy MBSs but they remain  small,   entanglement generation is possible only with a finite energy splitting of MBSs.

For asymmetric couplings at $\lambda = 0.7$, the situation is slightly different but with some similarities; see Fig.\,\ref{Fig4}(e,g) for $\omega = 0$ and $\omega \neq0$ in the MS setup. First of all, $R_{p}$, $E_{d}$, and $D$ oscillate with time; see Fig.\,\ref{Fig4}(e,g). As for the symmetric case, $C=0$ throughout the evolution at $\omega = 0$ [Fig.\,\ref{Fig4}(e)],  meaning that entanglement can  not be generated. This is because the eigenenergies in Eqs.\,(\ref{MS}) are given by $\Delta = |\lambda - 1|$, which simplifies the coefficients in Eq.\,(\ref{cofMSsep}) as $c_{1(4)}(t) = [\cos{(\lambda-1)t}\mp\cos{(\lambda+1)t}]/2$, 
$c_{2(3)}(t) = \pm i[\sin{(\lambda+1)t}\mp\sin{(\lambda-1)t}]/2$. This gives $|c_1(t)||c_4(t)| = |c_2(t)||c_3(t)|=|\sin^2{(\lambda-1)t}-\sin^2{(\lambda+1)t}|$. Therefore, the concurrence $C$, as defined in Eq.\,(\ref{NFSmaxintRpEdCD}), goes to zero throughout the evolution; see Fig.\,\ref{Fig4}(e).  However, at $\omega\neq0$, $C$ takes finite values and develops an oscillatory profile; see Fig.\,\ref{Fig4}(g). In relation to the discord, it develops   oscillations with different periodicities, acquiring a beating profile that oscillates faster for finite frequencies, as seen in Fig.\,\ref{Fig4}(e,g). We note that the complex behaviour of $D$ stems from the fact that it is determined by different entropies; see Appendix \ref{AppendixA1}. When it comes to the  entanglement dynamics in asymmetric couplings ($\lambda\neq 1$) with $\omega=0$, we find that $E_d=\sin^2(t)/2$ which is independent of $\lambda$ and has a constant period of $\pi$ as seen Fig.\,\ref{Fig4}(e). In contrast, in the same regime, we obtain the return probability to be $R_p=\cos^2(\lambda t)\cos^2(t)$, which clearly depends on two periods, $\pi/\lambda$ and $\pi$. Therefore, the system takes a long time to return to the initial state, beyond the scope of Fig.\,\ref{Fig4}(e). For clarification, we note that the first maxima of $R_{p}$ happens at $t=(10\pi)\,\hbar/\lambda_1$. Moving into the asymmetric couplings ($\lambda\neq 1$) and finite $\omega\neq 0$, as shown in Fig.\,\ref{Fig4}(g), the quantities $R_{p}$, $E_{d}$, $C$, and $D$ exhibit oscillatory dynamics without showing any special {entanglement characteristics, unlike their counterparts for a maximally entangled initial state discussed in Section \ref{secmaxentangled}.}

In relation to the entanglement measures for the NFS system, shown in Fig.\,\ref{Fig4}(b,d,f,h), they exhibit a beaviour that is roughly similar either at finite or zero energies and at symmetric or asymmetric couplings. Here, the concurrence we obtain is equal to $C= ||\bar{c}^\prime_2(t)||\bar{c}^\prime_3(t)||$, which goes to zero when either of the coefficients approaches zero.  The time at which $C=0$ is calculated as $t= (n\pi/{\sqrt{1+\lambda^2+\omega^2}})\, \hbar/\lambda_1$, where n is an integer.  At these points in time, the NFS acquires the state $\ket{110}$, which is a zero concurrence state ($C=0$). This can be verified from the results in Fig.\,\ref{Fig4}(b,d,f,h), where zero concurrence occurs between $t=2\,\hbar/\lambda_1$ and $t=3\,\hbar/\lambda_1$, depending on the particular value of $\lambda$ and $\omega$ during the evolution. 

 \begin{figure}[t]
  \includegraphics[width=0.49\columnwidth]{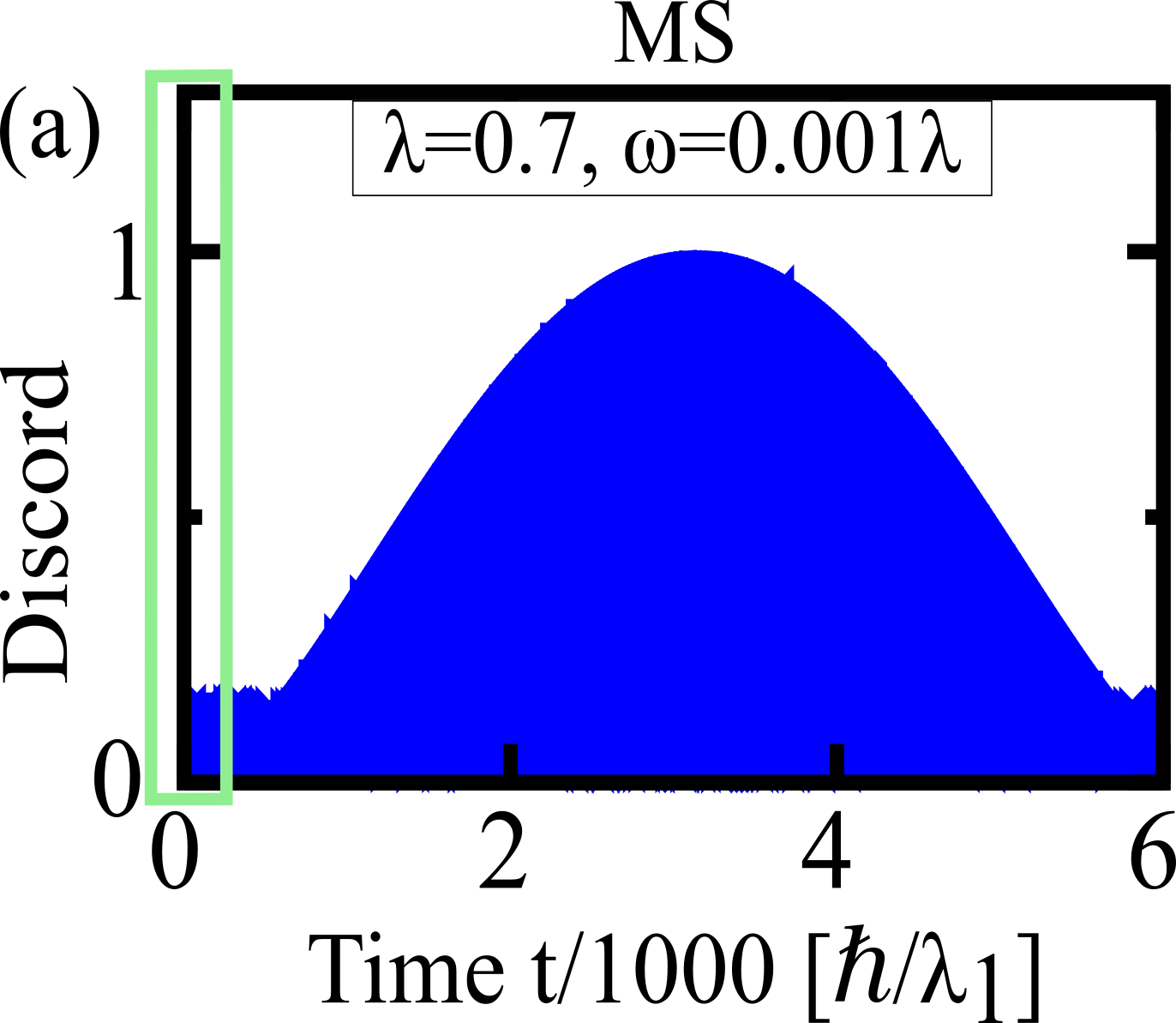}\hfill
  \includegraphics[width=0.49\columnwidth]{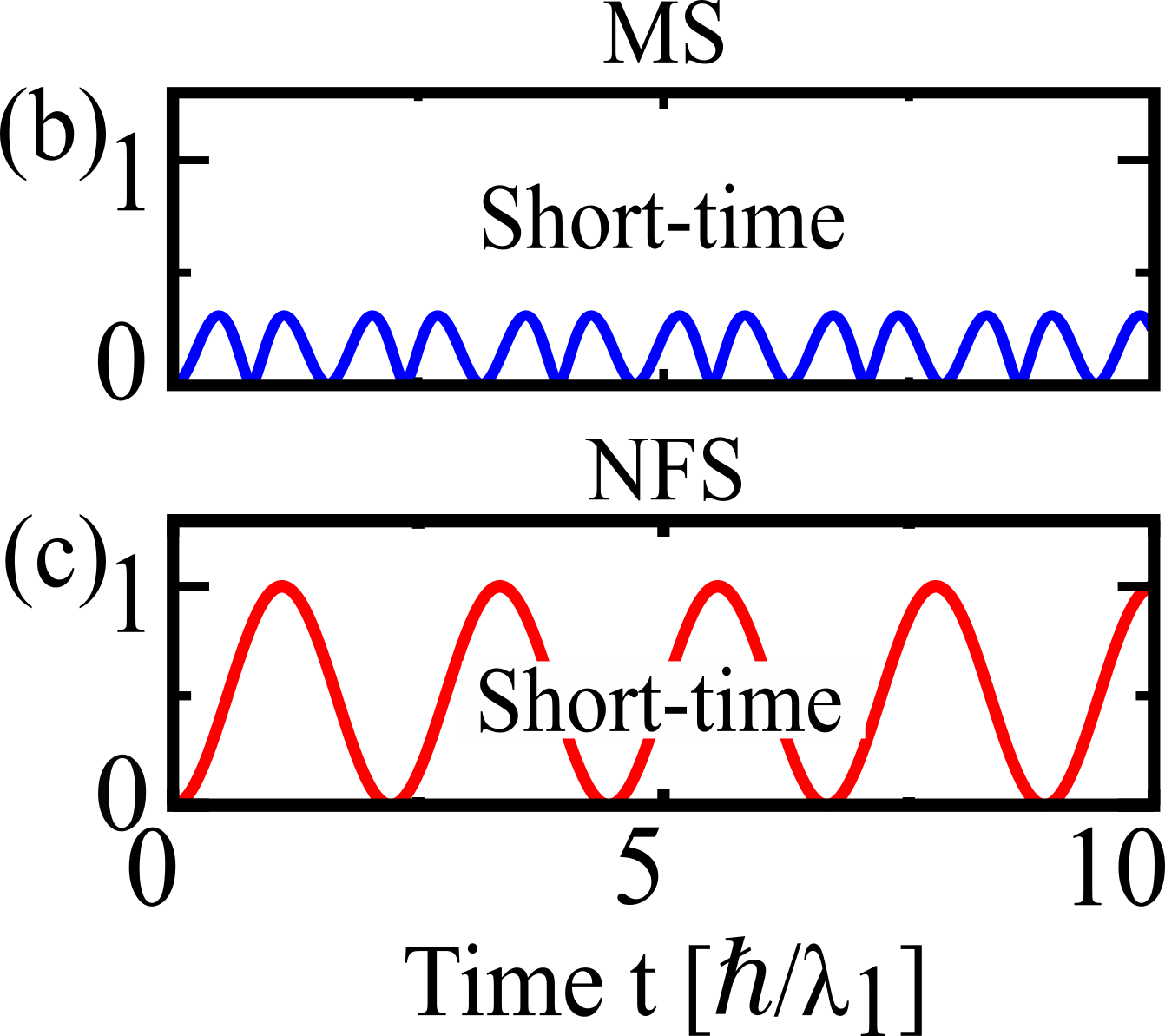}

  \caption{Time evolution of discord at $\lambda=0.7$ and $\omega = 0.001\lambda$ for a MS (a,b) and a NFS (c).  Panel (a) shows the long-time evolution, while (b,c) shows the short-time behaviour. The short times for the MS in (b) are indicated by the green box in (a). We note that, for these parameters, the long-time evolution of discord for  the  NFS has the same behaviour as the short-time counterpart in (c).}
  \label{Fig5}
\end{figure}

\subsection{Generating maximally entangled QDs induced by MBSs}
We have seen in the previous subsection that the time-evolution in the MS setup is sensitive to variations of  $\omega$. Of particular interest in this part is the behaviour of the quantum correlations at $\omega\sim0$, because this regime has well localized MBSs in the MS. Here we exploit the sensitivity of discord $D$ at $\omega\sim0$ to generate a maximally entangled state between QDs. The time evolution of $D$ at $\omega=0.001$ and asymmetric coupling $\lambda=0.7$  is presented in  Fig.\,\ref{Fig5}(a,b)  for the MS setup for long and short times, with Fig.\,\ref{Fig5}(b)  taken within the green box Fig.\,\ref{Fig5}(a). We contrast this behavior with the time evolution for the NFS at short times. The first observation is that $D$ exhibits rapid oscillations with an initial small amplitude of $0.1$ at short times [Fig.\,\ref{Fig5}(b)]. As time progresses,    $D$ increases and eventually reaches unity in the long-term evolution, see Fig.\,\ref{Fig5}(a). Having $D=1$ means that a maximally entangled stated of QDs is created, achieved only at a small energy splitting of MBSs. In contrast to the MS, the discord for the NFS system  in Fig.\,\ref{Fig5}(c) develops homogeneous oscillations at short times whose period and amplitudes do not change over time. We have verified that this behaviour in the NFS also remains in the long-time dynamics, making it very different to the MS.

\section{Concluding remarks}
\label{section5}
In conclusion, we have investigated the entanglement properties of Majorana bound states emerging in a topological superconductor that is coupled to two quantum dots. To contrast the impact of Majorana bound states, we have compared this Majorana system with an equivalent where the quantum dots are coupled via a normal fermion. In order to characterize entanglement in both systems, we have employed concurrence and discord. We also complemented this by studying the entanglement dynamics and return probability. We found that zero-energy Majorana bound states can transform initially maximally entangled quantum dots into a classical state, while maximally entangled states are notably achieved at finite Majorana energies, thus highlighting the impact of Majorana nonlocality. Remarkably, we have shown that the maximal entanglement between quantum dots at the beginning of the evolution can be completely shifted to a state between Majorana bound states and one of the quantum dots.

Furthermore, we found that entanglement and quantum correlations can be generated from an initially separable state in the Majorana system, with discord reliably quantifying quantum correlations for zero-energy Majorana bound states. We also showed that entanglement generation in the Majorana system is very sensitive to the nonlocality of the Majorana bound states, which, upon an appropriate control, can even induce maximally entangled states in the long-time dynamics.  
The inherent Majorana nonlocality originates distinct entanglement features in the Majorana  and normal fermion systems, thus highlighting the potential use of quantum correlations for distinguishing zero-energy states in Majorana devices and their utility for quantum information.
Our results can be in principle measured by reconstructing the density matrix via quantum state tomography \cite{altepeter2004tomography,doi:10.1126/science.1130886}, thus adding an alternative way for detecting Majorana states as well as offering another approach for generating maximally entanglement states beyond photon- \cite{PhysRevResearch.6.043029, Shadbolt_2011, Riedinger2018} and cavity-mediated systems \cite{PRXQuantum.5.020339, Vijayan2024}.

It is worth noting that our results assume that the parity of the system is fixed during the time evolution, being either even or odd. Under realistic conditions, however, this situation might be difficult to achieve because nonequilibrium quasiparticles or equilibrium nontopological subgap states  can poison  the computational subspace and lead to parity flips, see e.g., Refs.\,\cite{PhysRevLett.103.097002,rainis2012majorana,PhysRevB.95.235305,PhysRevB.97.125404,PhysRevLett.126.057702}. While quasiparticle poisoning is challenging to avoid in realistic setups, it has been  shown that   nonequilibrium quasiparticle poisoning can be  reduced  with Majorana zero modes because they act as robust quasiparticle traps, giving rise to  decoherence times longer than the time needed for qubit operations in moderately sized topological superconductors \cite{PhysRevLett.126.057702}. Moreover, the detrimental effect of quasiparticle poisoning due to nontopological subgap states,  very likely  due to  disorder-induced states which coexist with Majorana states \cite{PhysRevB.107.184519}, has been shown to be negligible at weak disorder \cite{PhysRevLett.126.057702}. Taking these ideas into account and under similar considerations, we expect that quasiparticle poisoning can be mitigated in the systems we study. We, however, anticipate that variations of the system's parity  might induce a change in the periodicity of the correlations dynamics, which can serve as a signature of  quasiparticle poisoning in our case. 


\section{Acknowledgements}
We thank J. C. Abadillo-Uriel, M. Benito,  V. A. Mousolou, and E.  Sj\"{o}qvist 
for insightful discussions. We acknowledge financial support from the Swedish Research Council  (Vetenskapsr\aa det Grant No.~2021-04121) and from  the Carl Trygger’s Foundation (Grant No. 22: 2093).


\appendix
\section{Calculation of quantum discord}
\label{AppendixA}
In this Appendix, we outline the procedure for computing quantum discord \cite{Vimal2} as defined in the main text for both MS and NFS. Here, for the comparative study of concurrence, we focus on the quantum discord between the two QDs, which can be calculated using the two-qubit reduced density matrix $\rho_d$ defined in Eq.\,(\ref{RDMX}) for both systems. We provide the procedure to compute discord for MS first, followed by that for NFS in the subsequent subsection. 

\subsection{Majorana system}
\label{AppendixA1}
For initially maximally entangled QDs in the MS, the matrix $\rho_d$ in Eq.\,(\ref{RDMX}) is computed by tracing over the nonlocal fermion from the state described in Eq.\,(\ref{psit}). On the other hand, for the separable initial state of the MS written in Eq.\,(\ref{separablest}), the same can be calculated from the state formulated by Eq.\,(\ref{cofMSsep}). In both scenarios, the resulting $\rho_d$ matrices represent the reduced density matrix of the two QDs. Therefore, the basis states and the form of matrices remain unchanged. Subsequently, for a better representation, the first QD is denoted as subsystem A and the second QD as subsystem B. Therefore, $\rho_d$ for the two QDs will be denoted as $\rho_{AB}$ and will have the same $X$-state representation as written in Eq.\,(\ref{RDMX}). For the discord calculation, we set the measurement basis of B as $\lbrace |\tilde{0}\rangle,|\tilde{1}\rangle \rbrace$, parametrized by $\theta$ and $\phi$, which can be transformed from the computational basis $\lbrace |0\rangle, |1\rangle\rbrace$ of B as
\begin{equation}
\label{TrBasis}
\begin{split}
&|\tilde{0}\rangle = \cos{\theta/2}|0\rangle +e^{i\phi}\sin{\theta/2}|1\rangle\,,\\
&|\tilde{1}\rangle = \sin{\theta/2}|0\rangle -e^{i\phi}\cos{\theta/2}|1\rangle\,,
\end{split}
\end{equation}
where the parameters $\theta \in [0,\pi]$ and $\phi\in[0,2\pi]$. In the new basis states $\lbrace |\tilde{k}\rangle \rbrace$ =$\lbrace |\tilde{0}\rangle, |\tilde{1}\rangle \rbrace$, we compute the marginal conditional density matrices $\rho_{A|B_{\tilde{k}}}$ to obtain the conditional entropy in Eq.\,(\ref{EqC}) as
\begin{equation}
\rho_{A|B_{\tilde{k}}} = \frac{1}{p_{\tilde{k}}}Tr_{B}|\tilde{k}\rangle\langle\tilde{k}|\rho_{AB}, \quad p_{\tilde{k}} = Tr_{AB}|\tilde{k}\rangle\langle\tilde{k}|\rho_{AB}\,,
\end{equation}
where the expressions for the probabilities $p_{\tilde{k}}$ are obtained using the element of $\rho_{AB}$, borrowed from Eq.\,(\ref{RDMX}) as
\begin{equation}
\label{probC}
\begin{split}
&p_{\tilde{0}} = (u+ w_2)\cos^2(\theta/2) +(v+w_1)\sin^2(\theta/2),\\
&p_{\tilde{1}} =  (u+ w_2)\sin^2(\theta/2) +(v+w_1)\cos^2(\theta/2)\,.
\end{split}
\end{equation}
Therefore, the eigenvalues of $\rho_{A|B_{\tilde{k}}}$ in terms of the above probabilities can be calculated as
\begin{equation}
\label{eigenC}
\begin{split}
 &\lambda_{\pm} = \frac{1}{2p_{\tilde{0}}}(p_{\tilde{0}}\pm \sqrt{b_{\tilde{0}}^2 + 4|z|^2}), \quad \text{for}\quad \rho_{A|B_{\tilde{0}}},\\
&\lambda_{\pm}' = \frac{1}{2p_{\tilde{1}}}(p_{\tilde{1}}\pm \sqrt{b_{\tilde{1}}^2 + 4|z|^2}),\quad \text{for}\quad \rho_{A|B_{\tilde{1}}}\,, 
\end{split}
\end{equation}
where, $z=\frac{1}{2}\sin{\theta}(e^{i\phi}x+e^{-i\phi}y) $ and
\begin{equation}
\label{btildeC}
\begin{split}
 &b_{\tilde{0}}=  (u-w_2)\cos^2(\theta/2) +(w_1-v)\sin^2(\theta/2), \\
 &b_{\tilde{1}}=  (u-w_2)\sin^2(\theta/2) +(w_1-v)\cos^2(\theta/2)\,.
\end{split}
\end{equation}
Using all the expressions from Eq.\,(\ref{probC}-\ref{btildeC}), the conditional entropy $C_{\theta,\phi}(\rho_{A|B_{\tilde{k}}})$ can be written by the weighted sum of marginal conditional entropies as 
\begin{equation}
\label{condent}
C_{\theta,\phi}(\rho_{A|B}) = p_{\tilde{0}}S(\rho_{A|B_{\tilde{0}}}) + p_{\tilde{1}}S(\rho_{A|B_{\tilde{1}}})\,,
\end{equation}
where the marginal entropies are computed using the eigenvalues written in Eq.\,(\ref{eigenC}) as follows
\begin{equation}
\label{margiEnt}
\begin{split}
&S(\rho_{A|B_{\tilde{0}}}) = -\lambda_+\log_2\lambda_+ -\lambda_-\log_2\lambda_- \,, \\
&S(\rho_{A|B_{\tilde{1}}}) = -\lambda_+^{\prime}\log_2\lambda_+^{\prime} -\lambda_-^{\prime}\log_2\lambda_-^{\prime}\, .
\end{split}
\end{equation}
Now, we can compute the quantum discord of the two QDs as defined in Eq.\,(\ref{discordQ}) as
\begin{equation}
\label{discoQDs}
D_{AB} = \min_{(\theta, \phi)} C_{\theta, \phi}(\rho_{A|B})- S(\rho_{AB})+S(\rho_B),
\end{equation}
where the conditional entropy $C_{\theta,\phi}(\rho_{A|B})$ is given by Eq.\,(\ref{condent}) and can be minimized computationally over the angles $\theta$ and $\phi$. The composite entropy $S(\rho_{AB})$ is calculated using the eigenvalues of the $\rho_{AB}$ borrowed from Eq.\,(\ref{RDMX}), which are $(u+v \pm \sqrt{(u-v)^2+4|y|^2})/2$ and $(w_1+w_2 \pm \sqrt{(w_1-w_2)^2+4|x|^2)/2}$. The last term, $S(\rho_{B})$, represents the entropy of the second QD. Here, the matrix $\rho_B$ is obtained from $\rho_{AB}$ by tracing over A. The entropy $S(\rho_{B})$ is then calculated using the eigenvalues $u+w_2$ and $v+w_1$ of $\rho_B$. Therefore, Eq.\,(\ref{discoQDs}) can be used to calculate the discord in Eq.\,(\ref{discordQ}) for the MS with different initial states written in Eq.\,(\ref{psit}) and Eq.\,(\ref{cofMSsep}) by first computing $\rho_d$s as described in Eq.\,(\ref{RDMX}) and following the steps to Eq.\,(\ref{discoQDs}) of this section. These calculations have been utilized in Sec. \ref{secmaxentangled} and Sec. \ref{secseprableinitialstate}.

\subsection{Normal fermion system}
\label{AppendixA2}
The computation of discord for NFS follows the same steps as for MS. However, for NFS, the reduced density matrix $\rho_d$ for the two QDs is obtained by tracing over the normal fermion, as described in Eq.\,(\ref{RDM}), from the state given in Eq.\,(\ref{NFSPSI}) for the maximally entangled initial state. Alternatively, for the separable initial state, $\rho_d$ can be calculated from the state in Eq.\,(\ref{cofNFSsep}). Then, similar to the MS in previous subsection, we use $\rho_{AB}$ for the reduced density matrices of the two QDs and follow the steps from equations (\ref{TrBasis}) to (\ref{discoQDs}) to compute all the relevant quantities. The minimization of the conditional entropy in Eq.\,(\ref{discoQDs}) is carried out numerically. Therefore, we directly refer to Eq.\,(\ref{discoQDs}) for computing the discord between the two QDs in NFS. The difference in discord from MS to NFS lies in the initial states used: Eq.\,(\ref{NFSPSI}) and Eq.\,(\ref{cofNFSsep}) for the maximally entangled and separable initial states of NFS, respectively. We utilize these calculations in Sections \ref{secmaxentangled} and \ref{secseprableinitialstate} for NFS.

\section{Entanglement measures for the odd sector}
\label{oddsector}
The dynamics of the Hamiltonian in Eq.\,(\ref{MajoranaH}) can be explored in the odd sector of excitation, following the step-by-step calculation of the even sector in Sec. \ref{Models}. In this Appendix, we provide an explicit calculation of the odd sector Hamiltonian and its state dynamics to highlight the differences observed in this sector. The Hamiltonian matrix, $H^o$, in the odd (o) sector basis states, $\ket{001}$, $\ket{010}$, $\ket{100}$, and $\ket{111}$ can be written as
\begin{equation}
\label{oH}
H^{o}= \begin{pmatrix}
\omega &  \lambda &  1 & 0	\\
\lambda  & -\omega & 0 & 1 \\
1 & 0 & -\omega &  \lambda \\
0 & 1  & \lambda  & \omega 
\end{pmatrix}\,,
\end{equation} 
\begin{figure*}
\centering
\includegraphics[width=0.96\textwidth, height=0.25\textwidth]{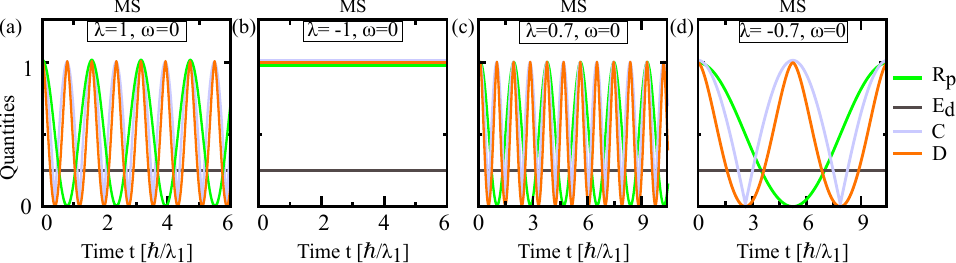}
\caption{Odd sector time evolution of concurrence, discord, return probability, and entanglement dynamics for an initially maximally entangled state of QDs in the MS. (a, b) Dynamics for $\lambda = 1$ and $\lambda = -1$ at $\omega = 0$. (c, d) Dynamics for $\lambda = 0.7$ and $\lambda = -0.7$ at $\omega = 0$. Negative values of $\lambda$ and $\omega$ can invert the dynamics, switching between odd and even sectors.}
\label{Fig6}
\end{figure*}
where the parameters $\lambda$ is the ration of couplings of MBSs and QDs and $\omega$ is the energy associated with the overlap of the MBSs wave functions. The matrix elements equal to 1 correspond to $\lambda_1 = 1$ (see Fig.\,\ref{Fig1}). This Hamiltonian matrix is similar in form to the one defined in the even sector in Eq.\,(\ref{Heven}), with the key difference being the sign of the coupling energy $\omega$ and $\lambda_1$. However, this sign change does not affect the eigenvalues, which are still given by the same equation as in Eq.\,(\ref{EnergyMS}). Therefore, the energy eigenvalues for the odd sector Hamiltonian can be written as
\begin{equation}
\label{oEnergyMS}
E^o_{i}=\pm\sqrt{(\lambda\pm1)^{2}+\omega^{2}}\,,
\end{equation}
where energy eigenvalues appear in pairs as  $E^o_{1(2)}=\mp\Delta$ and $E^o_{3(4)}=\mp\Delta_{1}$, with $\Delta = \sqrt{(\lambda-1)^2+\omega^2}$ and $\Delta_1 =\sqrt{(\lambda+1)^2+\omega^2}$. However, the eigenvectors differ from Eq.\,(\ref{MS}) in terms of local phases of different configurations. The eigenvectors $\ket{E^o_j}$ for the paired eigenvalues in Eq.\,(\ref{oEnergyMS}) can be written as un-normalized column vectors in the order of its corresponding eigenvalues as
\begin{equation}
\label{oMS}
EV^o = (\ket{E^o_1}|\ket{E^o_2}|\ket{E^o_3}|\ket{E^o_4})\,,
\end{equation}
where $\ket{E^o_1} = (-1, [\omega+\Delta]/[\lambda-1], -[\omega+\Delta]/[\lambda-1], 1)^T$ and T is the transpose operation. The eigenvector $\ket{E^o_2}=\ket{E^o_{1}(\Delta\rightarrow-\Delta)}$, and $\ket{E^o_3} = (1, -[\omega+\Delta_1]/[\lambda+1], -[\omega+\Delta_1]/[\lambda+1], 1)^T$, while $\ket{E^o_{4}}=\ket{E^o_{3}(\Delta_1\rightarrow -\Delta_{1})}$.
Similar to the even sector, all eigenvalues in Eq.\,(\ref{oEnergyMS}) depend on $\lambda$ and $\omega$, while the corresponding eigenstates in Eq.\,(\ref{oMS}) include contributions from all configurations of the odd states listed in Eq.\,(\ref{oH}). The redistribution of the local phases within the eigenstates of the odd sector differs from that in the even sector, as shown in Eq.\,(\ref{MS}). This difference affects the dynamics of entanglement measures and other dynamical quantities. Therefore, we provide calculations of the dynamical state functions in both general and specific cases to emphasize the similarities and differences in the dynamics between the two sectors.

\subsection{Evolution of a maximally entangled state}
In this section, we consider an initial state of maximally entangled QDs in the odd sector. For this purpose, we choose the state in the form of
\begin{equation}
\label{ois}
 |\psi^o(0)\rangle = \frac{|001\rangle + |111\rangle}{\sqrt{2}}\,.   
\end{equation}
In this initial state, the two QDs share the same form of initial entanglement as considered in Eq.\,(\ref{initialstate1}). To preserve the odd parity in the system, the nonlocal fermion formed from the MBSs is in the state $\ket{1}$. Subsequently, we allow the system to evolve with arbitrary values of the Hamiltonian parameters to compute all dynamical quantities. For this initial state, the system's dynamics can be expressed as
\begin{equation}
\label{opsitdef}
\begin{split}
    \ket{\psi^o(t)} &= e^{-iE^o_3t}\ket{E^o_3}\langle E^o_3|\psi^o(0)\rangle \\ 
    &+ e^{-iE^o_4t}\ket{E^o_4}\langle E^o_4|\psi^o(0)\rangle\,,
\end{split}
\end{equation}
where $|\psi^o(0)\rangle$ is the initial state defined in Eq.\,(\ref{ois}), and $|E^o_{3(4)}\rangle$ are given in Eq.\,(\ref{oMS}). In this dynamics, the system accesses only two eigenstates, $|E^o_{3(4)}\rangle$, as the other two states are orthogonal to the initial state. The expression in Eq.\,(\ref{opsitdef}) can be simplified in a similar form to the one given in Eq.\,(\ref{psit}). We can write
\begin{equation}
 \label{opsit}
 \begin{split}
\ket{\psi^o(t)} &= \frac{1}{\sqrt{2}}
\ket{0}( \eta^\prime\ket{01} + \chi^\prime\ket{10})\\
&+
\frac{1}{\sqrt{2}}\ket{1}(\eta^\prime\ket{11} + \chi^\prime\ket{00})\,,
\end{split}
\end{equation}
where $\chi^\prime$ and $\eta^\prime$ are time-dependent and given by
\begin{equation}
\label{etachip}
\begin{split}
\eta^\prime(t)&=\frac{1}{\Delta_1}(\Delta_1\cos{\Delta_1 t} - i \omega\sin{\Delta_1 t})\,,\\
\chi^\prime(t)&= \frac{-i}{\Delta_1}(\lambda+1)\sin{\Delta_1 t}\,.
\end{split}
\end{equation}
The expression in Eq.\,(\ref{opsit}) describes the time-evolved state of the system in the odd sector and has a similar form to that in Eq.\,(\ref{psit}) for the even sector. However, the expressions for $\eta^\prime$ and $\chi^\prime$ differ from those in the even sector. These coefficients involve the eigenvalue $\Delta_1$ and the term $\lambda+1$, rather than $\Delta$ and $\lambda-1$ as in Eq.\,(\ref{etachi}) for the even sector. We present the results in Fig.\,\ref{Fig6} for completely localized MBSs ($\omega=0$), considering both symmetric and asymmetric couplings to the QDs.
In Fig.\,\ref{Fig6}(a), all quantities exhibit periodic behavior except for the entanglement dynamics ($E_d$), which remains constant throughout. At $\omega=0$, $E_d$ can be expressed as $E_d = |e^{i(\Delta_1 t)}|^2/4$, which has a constant magnitude of 0.25. In contrast, concurrence and discord show the system transitioning from maximally entangled QDs ($C=D=1$) to separable states ($C=D=0$). The return probability is also periodic, with a periodicity twice that of concurrence. This can be explained by the simple expressions $C = \cos(4t)$ and $R_p = \cos^2(2t)$. The zeros of discord can be analyzed in a manner similar to the approach used in the even sector. Furthermore, in the case of asymmetric couplings of MBSs to QDs, we present all quantities in Fig.\,\ref{Fig6}(c). The dynamics are similar to those observed for symmetric coupling ($\lambda=1$), except that the periodicity changes due to a shift in the contributing eigenenergy values of the system's Hamiltonian.

\subsection{Comparing the dynamics of the  odd sector correlations with their even sector counterparts}
In Fig. \ref{Fig2}(a), we have shown that the even sector dynamics of the system show no evolution of entanglement measures and the state dynamical functions at $\omega= 0$ and $\lambda=1$. This feature can be retrieved in the odd sector as well by tuning the parameters carefully. The equivalent dynamics can be obtained by applying the transformations $\omega \rightarrow -\omega$ and $\lambda \rightarrow -\lambda$. Changing the sign of $\lambda$ indirectly changes the sign of $\lambda_1$, thus retrieving the Hamiltonian matrix in Eq.\,(\ref{oH}) of the even sector, as given in Eq.\,(\ref{Heven}). With the transformed parameters, the eigenstates of Eq.\,(\ref{oMS}) simplify, and we obtain the only contributing eigenvector of the form $\ket{E^o_3} = (|001\rangle + |111\rangle)/\sqrt{2}$, which is the same as the initial state $|\psi^o(0)\rangle$. The remaining eigenvectors are orthogonal to the initial state and thus do not contribute to the dynamics. Furthermore, since the eigenvector $E^o_{3}$ has zero energy, the dynamics of the initial state in Eq.\,(\ref{ois}) do not evolve at all, according to Eq.\,(\ref{timeevolved}). Consequently, we obtain the same results as shown in Fig.\,\ref{Fig2}(a), which are also displayed in Fig.\,\ref{Fig6}(b) at $\lambda = -1$ and $\omega = 0$ for the odd sector.

However, in the case of asymmetric couplings ($\lambda \neq 1$), it is important to note that the coefficients $\eta^\prime$ and $\chi^\prime$ are interchanged with their corresponding configurations compared to the results in Eq.\,(\ref{psit}) for the even sector. This interchange occurs primarily to preserve the odd parity of the system. Therefore, the dynamics remain the same as shown in Fig.\,\ref{Fig6}(d) at $\lambda = -0.7$ and $\omega = 0$. Similar changes are expected to occur for finite $\omega$ values, which we do not show here. Consequently, we conclude that the measurable quantities exhibit different dynamics for $\omega = 0$ and $\omega \neq 0$, similar to the even sector, but with differences in periodicity. More importantly, the
overall characteristics of the dynamics in both sectors remain consistent and can be analyzed independently, as dictated by the odd-even symmetry of the system.


\bibliography{biblio}

\end{document}